\documentstyle[11pt]{article}
\normalbaselines
\begin{document}

\newcommand{\m}[1]{\mbox{{\bf #1}}}
\newcommand{\Cl}{\mbox{C$_{\ell}$}}
\newcommand{\Al}{\mbox{A$_{\ell}$}}
\newcommand{\Ali}{\mbox{A$_{\ell}^{(1)}$}}
\newcommand{\Cli}{\mbox{C$_{\ell}^{(1)}$}}
\newcommand{\Bl}{\mbox{B$_{\ell}$}}
\newcommand{\Bli}{\mbox{B$_{\ell}^{(1)}$}}
\newcommand{\W}{\mbox{${\cal W}^{(1)}$}}
\newcommand{\T}{\mbox{$\tau^{\circ}$}}
\newcommand{\A}[1]{\mbox{$\alpha_{#1}^{\circ}$}}
\newcommand{\Ak}[1]{\mbox{$\alpha_{#1}$}}
\newcommand{\hk}[1]{\mbox{$h_{\alpha_{#1}}$}}
\newcommand{\D}[1]{\mbox{${\bf D}_{#1}$}}
\newcommand{\Dz}[1]{\mbox{${\bf D}^{\circ}_{#1}$}}
\newcommand{\Di}[1]{\mbox{${\bf D}^{-1}_{#1}$}}
\newcommand{\G}[1]{\mbox{${\bf G}_{#1}$}}
\newcommand{\Gh}[1]{\mbox{$\hat{{\bf G}}_{#1}$}}
\newcommand{\Gi}[1]{\mbox{${\bf G}_{#1}^{-1}$}}
\newcommand{\Ghi}[1]{\mbox{$\hat{{\bf G}}_{#1}^{-1}$}}
\newcommand{\HA}[1]{\mbox{${\bf H}_{#1}$}}
\newcommand{\Hh}[1]{\mbox{$\hat{{\bf H}}_{#1}$}}
\newcommand{\Hi}[1]{\mbox{${\bf H}_{#1}^{-1}$}}
\newcommand{\Hhi}[1]{\mbox{$\hat{{\bf H}}_{#1}^{-1}$}}
\newcommand{\E}[1]{\mbox{${\bf E}_{#1}$}}
\newcommand{\F}[1]{\mbox{${\bf F}_{#1}$}}
\newcommand{\Eh}[1]{\mbox{$\hat{{\bf E}}_{#1}$}}
\newcommand{\Ei}[1]{\mbox{${\bf E}_{#1}^{-1}$}}
\newcommand{\Fi}[1]{\mbox{${\bf F}_{#1}^{-1}$}}
\newcommand{\Ehi}[1]{\mbox{$\hat{{\bf E}}_{#1}^{-1}$}}
\newcommand{\LA}[1]{\mbox{${\bf L}_{#1}$}}
\newcommand{\Lt}[1]{\mbox{$\tilde{{\bf L}}_{#1}$}}
\newcommand{\Lp}[1]{\mbox{${\bf L}'_{#1}$}}
\newcommand{\LL}[1]{\mbox{${\bf L}^{\circ}_{#1}$}}
\newcommand{\LLt}[1]{\mbox{$\tilde{{\bf L}}^{\circ}_{#1}$}}
\newcommand{\Ltp}[1]{\mbox{$\tilde{{\bf L}}'_{#1}$}}
\newcommand{\Lpt}[1]{\Ltp{#1}}
\newcommand{\M}[1]{\mbox{${\bf M}_{#1}$}}
\newcommand{\Mt}[1]{\mbox{$\tilde{{\bf M}}_{#1}$}}
\newcommand{\Mti}[1]{\mbox{$\tilde{{\bf M}}_{#1}^{-1}$}}
\newcommand{\Mp}[1]{\mbox{${\bf M}'_{#1}$}}
\newcommand{\Mtp}[1]{\mbox{$\tilde{{\bf M}}'_{#1}$}}
\newcommand{\Mtpi}[1]{\mbox{$\tilde{{\bf M}}_{#1}^{'-1}$}}
\newcommand{\N}[1]{\mbox{${\bf N}_{#1}$}}
\newcommand{\Nt}[1]{\mbox{$\tilde{{\bf N}}_{#1}$}}
\newcommand{\Nti}[1]{\mbox{$\tilde{{\bf N}}^{-1}_{#1}$}}
\newcommand{\Np}[1]{\mbox{${\bf N}'_{#1}$}}
\newcommand{\Ntp}[1]{\mbox{$\tilde{{\bf N}}'_{#1}$}}
\newcommand{\Ntpi}[1]{\mbox{$\tilde{{\bf N}}_{#1}^{'-1}$}}
\newcommand{\zo}{\mbox{${\bf 0}$}}
\newcommand{\el}{\mbox{$\ell$}}
\newcommand{\dsum}{\mbox{dsum}}
\newcommand{\diag}{\mbox{diag}}
\newcommand{\offdiag}{\mbox{offdiag}}
\newcommand{\WW}[1]{\mbox{${\bf W}_{#1}$}}
\newcommand{\V}[1]{\mbox{${\bf V}_{#1}(t)$}}
\newcommand{\Vp}[1]{\mbox{${\bf V}_{#1}^{'}(t)$}}
\newcommand{\lt}[1]{\mbox{$\lambda_{#1}t^{\mu_{#1}}$}}
\newcommand{\lti}[1]{\mbox{$\lambda_{#1}^{-1}t^{-\mu_{#1}}$}}
\newcommand{\St}{\mbox{${\bf S}(t)$}}
\newcommand{\Sti}{\mbox{${\bf S}(t)^{-1}$}}
\newcommand{\Stii}{\mbox{${\bf S}(t^{-1})^{-1}$}}
\newcommand{\Ut}{\mbox{${\bf U}(t)$}}
\newcommand{\Uti}{\mbox{${\bf U}(t)^{-1}$}}
\newcommand{\Utm}{\mbox{${\bf U}(-t)$}}
\newcommand{\Upt}{\mbox{${\bf U}'(t)$}}
\newcommand{\Uppt}{\mbox{${\bf U}''(t)$}}
\newcommand{\Utp}{\mbox{${\bf U}'(t)$}}
\newcommand{\I}[1]{\mbox{${\bf 1}_{#1}$}}
\newcommand{\J}{\mbox{${\bf J}$}}
\newcommand{\bref}[1]{\mbox{(\ref{#1})}}
\newcommand{\arf}{\mbox{$\frac{1}{2}$}}
\newcommand{\SJS}{\mbox{$\tilde{{\bf S}}{\bf JS}$}}
\newcommand{\StJSt}{\mbox{$\tilde{{\bf S}}(t){\bf J}\St$}}
\newcommand{\lbd}{\lambda}
\newcommand{\sgm}[1]{\mbox{$\sigma_{#1}$}}
\newcommand{\wo}[1]{\mbox{$\rho_{#1}$}}
\newcommand{\Smti}{\mbox{${\bf S}(-t)^{-1}$}}
\newcommand{\Smt}{\mbox{${\bf S}(-t)$}}
\newcommand{\iE}{\mbox{{(E)}}}
\newcommand{\iF}[1]{\mbox{{(F)}$^{(#1)}$}}
\newcommand{\iG}[1]{\mbox{{(G)}$^{(#1)}$}}
\newcommand{\iH}{\mbox{{(H)}}}
\newcommand{\offsum}{\mbox{offsum}}
\newcommand{\iA}[1]{\mbox{{(A)}$^{(#1)}$}}
\newcommand{\iB}{\mbox{{(B)}}}
\newcommand{\iC}{\mbox{{(C)}}}
\newcommand{\iD}{\mbox{{(D)}}}
\newcommand{\elbt}{\mbox{$[\frac{\ell}{2}]$}}
\newcommand{\kwo}{\mbox{$\frac{1}{4}$}}
\newcommand{\np}{\mbox{$n_{+}$}}
\newcommand{\nm}{\mbox{$n_{m}$}}
\newcommand{\npt}{\mbox{$n_{+}^{t}$}}
\newcommand{\nmt}{\mbox{$n_{-}^{t}$}}

\bibliographystyle{unsrt}

\vbox{\vspace{38mm}}
\begin{center}
{\LARGE \bf Conjugacy Classes of Involutive Automorphisms\\[2mm]
of the $C_{\ell}^{(1)}$ Affine Kac-Moody Algebras}\\[5mm]

S.P.Clarke\\{\it Department of Physics and Astronomy, University of
St.Andrews,
\\North Haugh, St.Andrews, Fife, KY16 9SS, Scotland, U.K.}\\[3mm]
J.F.Cornwell\\{\it Department of Physics and Astronomy, University of
St.Andrews,
\\North Haugh, St.Andrews, Fife, KY16 9SS, Scotland, U.K.
\\(e-mail: J.F.Cornwell@St-Andrews.AC.UK)}\\[5mm]
(Submitted 6 June 1994)\\[5mm] \end{center}

\begin{abstract} The conjugacy classes of the involutive automorphisms of the
affine Kac-Moody
algebras \Cli\ for $\ell\geq 2$ are determined using the matrix formulation
of automorphisms of an affine Kac-Moody algebra.  PACS:
2.20.+b, 03.30.+p, 11.17.+y, 12.40Aa.
\end{abstract}

\begin{section}{Introduction}

The significant role played by the automorphism groups of semi-simple Lie
algebras is well known.In
particular, the study of the involutive automorphisms of complex semi-simple
Lie
algebras  by
Gantmacher \cite{Ga1} allowed Gantmacher \cite{Ga2} to obtain a very elegant
systematic
determination of all the simple real Lie algebras. As one can expect a similar
situation for the
affine Kac-Moody algebras, it is clearly important to determine the conjugacy
classes of the group
of automorphisms of the these algebras as well. Moreover, there are also
significant implications
for the Virasoro algebra and conformal field theory through the Sugawara
construction. The part
played by Kac-Moody automorphisms in this context has been discussed by
Bernard
\cite{Ber}, Walton \cite{Wal},  Bouwknegt \cite{Bou}, and Font \cite{Fon}.

	The structure of  the affine Kac-Moody algebras and their Weyl groups is
now very well
established.  (For reviews see Kac \cite{Ka1}, Goddard and Olive
\cite{Go1,Go2},
and Cornwell
\cite{Co0}).  Unless otherwise stated all the notations and conventions that
will be
employed in
the present paper are those of \cite{Co0}. In particular, quantities belonging
to the
simple complex Lie algebra ${\tilde{\cal{L}}}^{0}$ associated with an untwisted
affine Kac-Moody
algebra ${\tilde{\cal{L}}}$ are distinguished from the corresponding quantities
belonging to
${\tilde{\cal{L}}}$ by a superscript $0$, so that, for  example, $\alpha$ is
the
linear functional
on the Cartan subalgebra $\cal{H}$ of ${\tilde{\cal{L}}}$ that is the extension
of
the linear
functional ${\alpha}^{0}$ on the Cartan subalgebra
${\cal{H}}^{0}$ of ${\tilde{\cal{L}}}^{0}$.

     	   An automorphism of an affine Kac-Moody algebra ${\tilde{\cal{L}}}$
that
maps
every element of its Cartan subalgebra $\cal{H}$ into an element of  $\cal{H}$
is
called
a ``Cartan preserving automorphism''.  The study of this type of automorphism
was initiated by
Gorman {\em et al.} \cite{Gor} and extended further by Cornwell \cite{Co1}.
Although such automorphisms are very important, in that every conjugacy class
of
the
the automorphism group contains at least one Cartan preserving automorphism, it
is
necessary to go beyond such automorphisms.  The main reason for this is that it
is
possible for two Cartan preserving automorphisms to be conjugate members of the
group of all automorphisms of an affine Kac-Moody algebra, even though they are
not
conjugate within the subgroup of Cartan preserving automorphisms.  That is,
conjugacy of Cartan preserving automorphisms within the group of all
automorphisms
of an affine Kac-Moody algebra is often achieved via ``non-Cartan preserving
automorphisms''.

	To enable this problem to be tackled systematically,  a comprehensive
method of  dealing with {\em
all} the automorphisms of an untwisted affine Kac-Moody algebra based on a
matrix formulation  of
the untwisted affine Kac-Moody algebras was developed by Cornwell \cite{Co1}
(extending some
previous work on the corresponding ``derived subalgebra''
${\tilde{\cal{L}}}^{\prime}$ by
Levstein \cite{Lev}. It was shown in \cite{Co1} that in general there are  four
types of
automorphism  within this matrix formulation, which were called ``type 1a'',
``type
1b'', ``type
2a'', and ``type 2b''.  The explicit derivation of all of these types of
automorphism,
as well as the
investigation of identical automorphisms and of the identity automorphism, the
formulae for the
products of automorphisms, the conditions for an automorphism to be involutive,
the formulae for the
inverses of automorphisms, and the conjugacy conditions for automorphisms may
all be found
\cite{Co1}. The account given below just describes the parts that are essential
for
the present
analysis. Previous papers (Cornwell \cite{Co2,Co3,Co4}, Clarke and Cornwell
\cite{Cla}) have
discussed in detail the conjugacy classes of the involutive automorphisms of
the
algebras \Ali\ and
\Bli\ (for $\ell\geq 1$).

 The present  paper is devoted to a derivation of the conjugacy classes of the
involutive
automorphisms within the group of all automorphisms of \Cli, the analysis of
which presents a number
of new features. Its organization is as follows. In Section~\ref{sec-prelim}
the
essential facts
concerning the structure of \Cli\ and the relevant formulae of the matrix
formulation are very
briefly summarized, and a concise notation for certain matrices is presented.
The
transformations of
the Weyl group of
\Cl\ are given in Section~\ref{sec-Weyl}, and are used in the first subsection
of
Section~\ref{sec-listing} to deduce the matrices that generate the required
involutive automorphisms.
Some general results on the conjugacy problem for \Cli\ are obtained in the
second subsection of
Section~\ref{sec-listing}, and these are developed further for the three
specific
types of
automorphism that occur for \Cli\ in Section~\ref{sec-1a1},
Section~\ref{sec-1a2},
and
Section~\ref{sec-2a1}. The results are summarized in Section~\ref{sec-summ}.

\end{section}

\begin{section}{Preliminaries}
\label{sec-prelim}

\begin{subsection}{The structure of ${\tilde{\cal{L}}}$}

It is well known (\cite{Ka1,Go1,Go2,Co0}) that every complex untwisted affine
Kac-Moody algebra
${\tilde{\cal{L}}}$ can be constructed from its corresponding simple complex
Lie
algebra
${\tilde{\cal{L}}}^{0}$ in the following way.  ${\tilde{\cal{L}}}$ may be taken
to
have a general
element of the form
\begin{equation} \sum_{j=-\infty}^{\infty}\sum_{p=1}^{n^{0}} \;
\mu_{jp} \;t^{j} \otimes a_{p}^{0}\;+\;\mu_{c}c\;+\;\mu_{d}d ,\label{eq:B.1}
\end{equation} where $\mu_{jp}$,  $\mu_{c}$, and  $\mu_{d}$ are arbitrary
complex numbers, with only
a finite number of the $\mu_{jp}$ being non-zero. Here
$j$ takes any integer value, $a_{p}^{0}$ are the basis elements of
${\tilde{\cal{L}}}^{0}$ (where $p = 1,2, \ldots, n^{0}$, $n^{0}$ being the
order of
${\tilde{\cal{L}}}^{0}$, and where $t$ is a complex number).  The commutators
of
${\tilde{\cal{L}}}$ are given by
	\begin{equation} [ t^{j} \otimes a^{0} , t^{k} \otimes b^{0} ]  =   t^{j+k}
\otimes [ a^{0} ,
b^{0} ] +  j {\delta}^{j+k,0}  B^{0}(a^{0},b^{0}) c \label{eq:26.89}
\end{equation}
 (for all integers $j$ and $k$ and all $a^{0},b^{0} \in {\tilde{\cal{L}}}^{0}$,
where
the
commutators and  Killing form $B^{0}( \; , \; )$ of the right-hand side are
those of
${\tilde{\cal{L}}}^{0}$),
 \begin{equation} [ t^{j} \otimes a^{0} , c ]   =   0	 \label{eq:26.90}
\end{equation}
(for all
integers $j$ and all $a^{0} \in {\tilde{\cal{L}}}^{0}$, and where again the
commutator on the
right-hand side of (\ref{eq:26.89}) is that of ${\tilde{\cal{L}}}^{0}$),
	\begin{equation} [ d, t^{j} \otimes a^{0}  ]   =   j t^{j} \otimes a^{0} ,
\label{eq:26.91}
\end{equation} (for all integers $j$ and all $a^{0} \in
{\tilde{\cal{L}}}^{0}$), and
	\begin{equation}  [ d , c ]   =   0  .	\label{eq:26.92} \end{equation}
If $h_{{\alpha}_{k}^{0}}^{0}$ (for $k = 1,2, \ldots ,\ell$) are the Weyl basis
elements of the
Cartan subalgebra ${\cal{H}}^{0}$ of ${\tilde{\cal{L}}}^{0}$ corresponding to
the
simple roots
$\alpha_{k}^{0}$ (for $k = 1,2, \ldots ,\ell$) of ${\tilde{\cal{L}}}^{0}$, then
the
$\ell+2$ basis elements $c$, $d$, and $t^{0} \otimes h_{{\alpha}_{k}^{0}}^{0}$
(for
$k = 1,2, \ldots
,\ell$) constitute a basis for a Cartan subalgebra $\cal{H}$ of
${\tilde{\cal{L}}}$.
The basis of the
``derived subalgebra'' ${\tilde{\cal{L}}}^{\prime}$ of ${\tilde{\cal{L}}}$
consists of
all the
basis elements of ${\tilde{\cal{L}}}$ except $d$.

 Every linear functional, and, in particular, every root, ${\alpha}^{0}$ that
is
defined on
${\cal{H}}^{0}$ can be "extended" to give a linear functional
${\alpha}$ on $\cal{H}$ by the definitions
	\begin{equation}  {\alpha}(t^{0} \otimes h_{{\alpha}_{k}^{0}}^{0}) =
{\alpha}^{0}(h_{{\alpha}_{k}^{0}}^{0}) \; \mbox{(for $k = 1,2,
\ldots ,\ell$)}, {\alpha}(c)   =   0  , {\alpha}(d)  =   0 .
\label{eq:26.95a} \end{equation} Moreover, if $\delta$ is the linear functional
on
$\cal{H}$
defined by
\begin{equation}  {\delta}(t^{0} \otimes h_{{\alpha}_{k}^{0}}^{0}) =  0 \;
\mbox{(for $k = 1,2,
\ldots ,\ell$)} ,{\delta}(c)   =   0
,{\delta}(d)  =   1 ,\label{eq:26.96a} \end{equation}
 then, for every integer value of $j$, $t^{j} \otimes
e_{{\alpha}^{0}}^{0}$  corresponds to a root
$j{\delta}+{\alpha}$ of ${\tilde{\cal{L}}}$ and   $t^{j} \otimes
h_{{\alpha}_{k}^{0}}^{0}$
corresponds to a root $j{\delta}$ of ${\tilde{\cal{L}}}$ (for $k = 1,2, \ldots
,\ell$).
Moreover,
for $k = 1,2, \ldots ,\ell$ the simple roots $\alpha_{k}$ of
${\tilde{\cal{L}}}$
are just the extensions of the simple roots $\alpha_{k}^{0}$ of
${\tilde{\cal{L}}}^{0}$, and the
remaining simple root $\alpha_{0}$ of ${\tilde{\cal{L}}}$ is given by
$\alpha_{0} =
\delta -
\alpha_{H}$, where $\alpha_{H}$ is the extension of the highest root
$\alpha_{H}^{0}$ of
${\tilde{\cal{L}}}^{0}$.

\end{subsection}

\begin{subsection}{Matrix formulation of the automorphisms of
${\tilde{\cal{L}}}$}

   Let ${\bf \Gamma}$ be a faithful irreducible representation  of some
dimension
$d_{\bf \Gamma}$ of ${\tilde{\cal{L}}}^{0}$, and let $\gamma$ be the Dynkin
index of ${\bf
\Gamma}$.  Then the first term $\sum_{j=-
\infty}^{\infty}\sum_{p=1}^{n^{0}}\mu_{jp} \;t^{j}
\otimes a_{p}^{0}$ of the general element (\ref{eq:B.1}) of the affine
untwisted
Kac-Moody algebra ${\tilde{\cal{L}}}$ is represented by the $d_{\bf \Gamma}
\times d_{\bf \Gamma}$ matrix $\sum_{j=-\infty}^{\infty}\sum_{p=1}^{n^{0}}
\mu_{jp}t^{j}{\bf \Gamma}(a_{p}^{0})$.  A typical matrix of this form will be
denoted by ${\bf a}(t)$, i.e. \begin{equation}{\bf
a}(t)=\sum_{j=-\infty}^{\infty}\sum_{p=1}^{n^{0}} \; \mu_{jp} \;t^{j}{\bf
\Gamma}(a_{p}^{0}).	\label{eq:B.2} \end{equation} Clearly all the entries
of
${\bf a}(t)$ are Laurent polynomials in the complex variable $t$.  A typical
element of ${\tilde{\cal{L}}}$ can then be written as
\begin{equation}{\bf a}(t)\;+\;\mu_{c}c\;+\;\mu_{d}d.	\label{eq:B.3}
\end{equation} (Of course in {\em no} sense do the  + signs in (\ref{eq:B.3})
represent ordinary matrix addition).

Each of the  four types (1a, 1b, 2a, and 2b) of automorphism of
${\tilde{\cal{L}}}$
depends on the
following three quantities (although the dependence is different for the
different
types):
\begin{enumerate}
 \item a $d_{\bf \Gamma} \times d_{\bf \Gamma}$ matrix ${\bf U}(t)$, which is
assumed to be invertible and for which all the entries of ${\bf U}(t)$ and
${\bf
U}(t)^{-1}$ are assumed to be Laurent polynomials in the complex variable $t$;
 \item a non-zero complex parameter $u$;
 \item an arbitrary complex parameter $\xi$.
\end{enumerate}
It is convenient to exhibit these together as a triple in the form $\{{\bf
U}(t),u,\xi\}$.  As noted below, for \Cli\ the type 1b automorphisms and the
type
2b
automorphisms coincide with the type 1a and the type 2a automorphisms
respectively, so only the
actions of type 1a and type 2a automorphisms $\phi$ need be explicitly
displayed:
\begin{enumerate}
 \item  Actions on {\bf a}(t):
    \begin{enumerate}
    \item For type 1a automorphisms:
\begin{eqnarray}
 \phi({\bf a}(t)) & = &{\bf U}(t){\bf a}(ut){\bf U}(t)^{-1} \nonumber \\
& &  +\;\frac{1}{\gamma}{\rm Res}\{{\rm tr}({\bf
U}(t)^{-1}\;\frac{d{\bf U}(t)}{dt}\;{\bf a}(ut))\}c.\label{eq:B.134}
\end{eqnarray}
    \item For type 2a automorphisms:
\begin{eqnarray}
 \phi({\bf a}(t)) & = &{\bf U}(t){\bf a}(ut^{-1}){\bf U}(t)^{-1} \nonumber \\
& &  +\;\frac{1}{\gamma}{\rm Res}\{{\rm tr}({\bf
U}(t)^{-1}\;\frac{d{\bf U}(t)}{dt}\;{\bf a}(ut^{-1}))\}c.\label{eq:B.167}
\end{eqnarray}
\end{enumerate}
\item  Actions on $c$ and $d$:
\begin{equation}\phi(c) =\mu c,\label{eq:B.189}\end{equation}
and
\begin{equation}\phi(d) =\mu {\bf {\Phi}}({\bf U}(t))\;+\;\xi
c\;+\;\mu d,\label{eq:B.190a}\end{equation}
where ${\bf {\Phi}}({\bf U}(t))$ is the $d_{\bf \Gamma}
\times d_{\bf \Gamma}$ matrix that depends on ${\bf U}(t)$ according to the
definition
  \begin{equation}{\bf {\Phi}}({\bf U}(t)) =\{-t\frac{d{\bf
U}(t)}{dt}\;{\bf U}(t)^{-1}\;+\;\frac{1}{d_{\bf \Gamma}}{\rm tr}(t\frac{d{\bf
U}(t)}{dt}\;{\bf U}(t)^{-1}){\bf 1}\},\label{eq:B.190b}\end{equation}
and
\begin{equation} \mu = \left\{ \begin{array}{ll}
 1 & \mbox{for type 1a,} \\
 -1 & \mbox{for type 2a.} \end{array}  \right.
\label{eq:B.191} \end{equation}
\end{enumerate}

If the triples $\{ {\bf U}(t),u,\xi \}$ and $\{ {\bf U'}(t),u',\xi' \}$ specify
two automorphisms of the {\em same} type, then these automorphisms are {\em
identical} if and only if
\begin{equation}u' = u ,\; \xi' = \xi ,\label{eq:new17} \end{equation}
and there exists a non-zero complex number $\eta$ and an
integer $k$ such that
\begin{equation}{\bf U'}(t)= \eta t^{k}{\bf
U}(t) . \label{eq:new18} \end{equation}

A type 1a automorphism corresponding to the triple $\{ {\bf U}(t),u,{\xi} \}$
is
involutive if and only if the following three conditions are all satisfied:
 \begin{equation}{\bf U}(t)\;{\bf U}(ut)=\eta t^{k} {\bf 1} \label{eq:B.149}
\end{equation}
for some complex number $\eta$ and some integer $k$,
 \begin{equation}u^{2} =
1, \label{eq:B.150} \end{equation} and
\begin{equation}\xi = -\frac{1}{2\gamma}{\rm Res}\{{\rm
tr}({\bf U}(t)^{-1}\;\frac{d{\bf U}(t)}{dt}\;{\bf {\Phi}}({\bf
U}(ut)))\}. \label{eq:B.151} \end{equation}
Similarly a
type 2a automorphism corresponding to the triple $\{ {\bf U}(t),u,{\xi} \}$ is
involutive if and only if the following two conditions are all satisfied:
 \begin{equation}{\bf U}(t)\;{\bf U}(ut^{-1})=\eta t^{k} {\bf 1}
\label{eq:B.198} \end{equation}
for some complex number $\eta$ and some integer $k$,
and
\begin{equation}\xi = -\frac{1}{2\gamma}{\rm Res}\{{\rm
tr}({\bf U}(t)^{-1}\;\frac{d{\bf U}(t)}{dt}\;{\bf {\Phi}}({\bf
U}(ut^{-1})))\}. \label{eq:B.199} \end{equation}

   The necessary and sufficient conditions for the conjugacy of a pair of
automorphisms $\phi_{1}$ and $\phi_{2}$ of ${\tilde{\cal{L}}}$ corresponding to
the triples $\{ {\bf U}_{1}(t),u_{1},{\xi}_{1} \}$ and $\{ {\bf
U}_{2}(t),u_{2},{\xi}_{2} \}$ respectively via an automorphism $\phi$ of
${\tilde{\cal{L}}}$  corresponding to the triple $\{ {\bf S}(t),s,{\xi} \}$
will now be stated.  More precisely, these are the necessary and
sufficient conditions for the automorphism equality $\phi_{1}=\phi \circ
\phi_{2} \circ \phi^{-1}$ to hold.  In each case there three such conditions,
but only two will be exhibited explicitly. These are the conditions relating
the
matrix ${\bf U}_{1}(t)$ to the matrix ${\bf U}_{2}(t)$ and the parameter
$u_{1}$
to the parameter $u_{2}$.  It is possible to also relate the parameter
$\xi_{1}$
to the parameter $\xi_{2}$, but the resulting expressions will be omitted as
they are very complicated and will not be needed in the subsequent analysis. If
$\phi_{1}$ and
$\phi_{2}$ are conjugate then they must be of the same type.
\begin{enumerate}
\item  If $\phi_{1}$ and $\phi_{2}$ are two type 1a automorphisms and $\phi$
is a type 1a automorphism, the conditions are:
\begin{equation}\eta t^{k}{\bf U}_{1}(t) = {\bf S}(t)\;{\bf U}_{2}(st)\;{\bf
S}(u_{2}t)^{-1} , \label{eq:B.155} \end{equation}
where $\eta$ is any non-zero complex number and $k$ is any integer, and
\begin{equation} u_{1} = u_{2} . \label{eq:B.156} \end{equation}
\item  If $\phi_{1}$ and $\phi_{2}$ are two type 1a automorphisms and $\phi$
is a type 2a automorphism, the conditions are:
\begin{equation}\eta t^{k}{\bf U}_{1}(t) = {\bf S}(t)\;{\bf
U}_{2}(st^{-1})\;{\bf
S}(u_{2}^{-1}t)^{-1} , \label{eq:B.225} \end{equation}
where $\eta$ is any non-zero complex number and $k$ is any integer, and
\begin{equation} u_{1} = u_{2}^{-1} . \label{eq:B.226} \end{equation}
\item  If $\phi_{1}$ and $\phi_{2}$ are two type 2a automorphisms and $\phi$
is a type 1a automorphism, the conditions are:
\begin{equation}\eta t^{k}{\bf U}_{1}(t) = {\bf S}(t)\;{\bf
U}_{2}(st)\;{\bf S}(s^{-2}u_{2}t^{-1})^{-1} , \label{eq:B.202} \end{equation}
where $\eta$ is any non-zero complex number and $k$ is any integer, and
\begin{equation} u_{1} = s^{-2}u_{2} . \label{eq:B.203} \end{equation}
\item  If $\phi_{1}$ and $\phi_{2}$ are two type 2a automorphisms and $\phi$
is a type 2a automorphism, the conditions are:
\begin{equation}\eta t^{k}{\bf U}_{1}(t) = {\bf S}(t)\;{\bf
U}_{2}(st^{-1})\;{\bf S}(s^{2}u_{2}^{-1}t^{-1})^{-1} , \label{eq:B.208}
\end{equation} where $\eta$ is any  non-zero complex number and $k$ is any
integer, and
\begin{equation} u_{1} = s^{2}u_{2}^{-1} . \label{eq:B.209} \end{equation}
\end{enumerate}

Taken together, these results imply that in searching for the conjugacy classes
of
involutive
automorphisms of \Cli\ only three categories of class representatives need be
considered, namely:
\begin{enumerate}
\item type 1a involutive automorphisms with $u=1$,
\item type 1a involutive automorphisms with $u=-1$,
\item type 2a involutive automorphisms with $u=1$.
\end{enumerate}
\end{subsection}

\begin{subsection} {The affine Kac-Moody algebra \Cli}
The algebra \Cli\ has ($\ell+1$) simple roots \Ak{0},$\ldots$,\Ak{\el} and
generalized Cartan matrix {\bf A}, where
\[
{\bf A}=\left(\begin{array}{cccccccc}
2&-1&0&0&\cdots&0&0&0\\
-2&2&-1&0&\cdots&0&0&0\\
0&-1&2&-1&\cdots&0&0&0\\
0&0&-1&2&\cdots&0&0&0\\
\vdots&\vdots&\vdots&\vdots&\ddots&\vdots&\vdots&\vdots\\
0&0&0&0&\cdots&2&-1&0\\
0&0&0&0&\cdots&-1&2&-2\\
0&0&0&0&\cdots&0&-1&2\end{array}\right).
\]
The quantities $<\Ak{j},\Ak{k}>$,
are given by
\[ <\Ak{j},\Ak{k}>=B_{jk}, \]
where the matrix {\bf B} is
\[ {\bf B}=\frac{1}{4(\ell+1)}\left(\begin{array}{cccccccc}
4&-2&0&0&\cdots&0&0&0\\
-2&2&-1&0&\cdots&0&0&0\\
0&-1&2&-1&\cdots&0&0&0\\
0&0&-1&2&\cdots&0&0&0\\
\vdots&\vdots&\vdots&\vdots&\ddots&\vdots&\vdots&\vdots\\
0&0&0&0&\cdots&2&-1&0\\
0&0&0&0&\cdots&-1&2&-2\\
0&0&0&0&\cdots&0&-2&4\end{array}\right).\]
(For the matrices {\bf A} and {\bf B} the index set is taken to
be $\{0,1,\ldots,\ell\}$). As the associated simple complex Lie algebra for
\Cli\ is
the algebra
\Cl, it follows that $\alpha_{H} = 2(\Ak{1}+\cdots+\Ak{\ell-1})+\Ak{\ell}$, and
hence that
\[\begin{array}{lll}
\delta&=&\Ak{0}+2(\Ak{1}+\cdots+\Ak{\ell-1})+\Ak{\ell}\\
c&=&\hk{0}+2(\hk{1}+\cdots+\hk{\ell-1})+\hk{\ell}.\end{array}\]
As \Cl\ is the complexification of the real Lie algebra $sp(\ell)$, \Cl\ may be
taken to be the set
of $(2\ell\times 2\ell)$ complex matrices {\bf a} that satisfy
\begin{equation}\label{int5}\begin{array}{cc}
\tilde{{\bf a}}\J+\J{\bf a}={\bf 0}, &\mbox{where }\J=\left(\begin{array}{cc}

					\zo&\I{\ell}\\-
\I{\ell}&\zo\end{array}\right).\end{array}
\end{equation} If a matrix \Ut\ is to correspond to an automorphism of \Cli,
then
the mapping
\[ {\bf a}(t)\mapsto\Ut{\bf a}(ut)\Uti\] must stabilize the `matrix part' of
the
algebra. That is,
the image of ${\bf a}(t)$ must satisfy \bref{int5}. Schur's lemma implies that
this
condition is
satisfied if
\[ \tilde{{\bf U}}(t)\J\Ut=f(t)\J , \] where $f(t)$ is some function of $t$.
With the
assumption that
\Ut\ and \Uti\ have entries that are Laurent polynomials, then
\mbox{det(\Ut)}$=\alpha t^{\beta}$,
where $\alpha$ is some non-zero complex number and $\beta$ is some integer.

For \Cl\ the representation ${\bf \Gamma}$ of (\ref{eq:B.2}) may be defined as
follows. Let
\[ \A{j}=\left\{\begin{array}{ll}
\varepsilon_{j}-\varepsilon_{j+1},&j=1,2,\ldots,\ell-1\\
2\varepsilon_{\ell}&j=\ell,\end{array}\right.\]
so that the roots of \Cl\ may be expressed in the form
\[\begin{array}{ll}
(\varepsilon_{j}-\varepsilon_{k})&\mbox{ for }1\leq j<k\leq\ell\\
(\varepsilon_{j}+\varepsilon_{k})&\mbox{ for }1\leq j\leq
k\leq\ell.\end{array}\]
Then the matrices representing the basis elements of \Cl\ may be taken to be
\[\begin{array}{c}
{\bf {\Gamma}}(h_{\alpha_{j}})=\left\{\begin{array}{ll}
					\{4(\ell+1)\}^{-1}(\m{X}_{jj}-
\m{X}_{j+1,j+1})&\mbox{for }j=1,2,\ldots,\ell-1\\
					\{2(\ell+1)\}^{-1}\m{X}_{\ell,\ell}&\mbox{for
}j=\ell\end{array}\right.\\
\begin{array}{ll}
{\bf {\Gamma}}(e_{\varepsilon_{j}-\varepsilon_{k}})=\{4(\ell+1)\}^{-
\frac{1}{2}}\m{X}_{jk}&
\mbox{for }j<k;j,k=1,2,\ldots,\ell\\
{\bf
{\Gamma}}(e_{\varepsilon_{j}+\varepsilon_{k}})=\{4(\ell+1)(1+\delta_{jk})\}^{-
\frac{1}{2}}
\m{Y}_{jk}
&\mbox{for }j\leq k;j,k=1,2,\ldots,\ell,\end{array}\end{array}\]
where the matrices $\m{X}$ and $\m{Y}$ are defined below in (\ref{eq:X}). This
representation
${\bf
\Gamma}$ is equivalent to its contragredient representation, since
\[\tilde{{\bf {\Gamma}}}=-\J{\bf {\Gamma}}\J^{-1} ,\]
which implies that the type 1b automorphisms and the type 2b automorphisms
coincide with the type 1a
and the type 2a automorphisms respectively. The Dynkin index of this
representation is given by
\begin{equation} \gamma=\frac{1}{2(\ell+1)} . \end{equation}

\end{subsection}

\begin{subsection}{Concise notation for matrices}

It is necessary to introduce some notation here to make subsequent
analysis more concise. The expressions `dsum' and
`offsum' are analogous to the commonly-used `diag'.

\begin{enumerate}
\item The first of these indicates a direct sum. For example
\[ \dsum\{\m{a},\m{b},\ldots,\m{y},\m{z}\}=\left(\begin{array}{ccccc}
\m{a}&\zo&\cdots&\zo&\zo\\
\zo&\m{b}&\cdots&\zo&\zo\\
\vdots&\vdots&\ddots&\vdots&\vdots\\
\zo&\zo&\cdots&\m{y}&\zo\\
\zo&\zo&\cdots&\zo&\m{z} \end{array}\right),\]
where \m{a}, \m{b},$\ldots$, \m{y}, \m{z}\ are square matrices.

\item The expression `offsum' is similar to the  previous expression. Thus
\[\offsum\{\m{a},\m{b},\ldots,\m{y},\m{z}\}=\left(\begin{array}{ccccc}
\zo&\zo&\cdots&\zo&\m{a}\\
\zo&\zo&\cdots&\m{b}&\zo\\
\vdots&\vdots&\ddots&\vdots&\vdots\\
\zo&\m{y}&\cdots&\zo&\zo\\
\m{z}&\zo&\cdots&\zo&\zo \end{array}\right),\]
where \m{a}, \m{b},$\ldots$, \m{y}, \m{z} are all square matrices.

\item The ($\ell\times\ell$) matrices $\m{e}_{pq}$ (where $1\leq p,q\leq\ell$)
are defined by
\[(e_{pq})_{jk}=\delta_{jp}\delta_{kq} .\]
Then $\m{X}_{pq}$ and $\m{Y}_{pq}$ (where $1\leq p,q\leq\ell$)
may be defined by
\begin{equation}\begin{array}{cc}
\m{X}_{pq}=\dsum\{\m{e}_{pq},-\m{e}_{qp}\},&\m{Y}_{pq}=\offsum\{
(\m{e}_{pq}+\m{e}_{qp}),\zo\}.\end{array} \label{eq:X} \end{equation}

\item Some `general' matrices will now be defined. These are matrices which are
used to
typify whole collections of specific matrices. The `general' matrix \D{j,k}\
is defined to be the general $(k-j+1)\times(k-j+1)$ matrix that satisfies
\[\begin{array}{cc}
\D{j,k}=\diag\{1,\lt{j+1},\ldots,\lt{k}\}&\lbd_{q}^{2}=1\mbox{ (for }
j+1\leq q\leq k)\end{array},\]
while \Dz{j,k}\ is of the general form \D{j,k}\ but has the additional
constraint
that $\mu_{j+1}=\cdots=\mu_{k}=0$. (For all of the general matrices it is
assumed that $\lbd_{q}$ is a non-zero complex number (for $1\leq q\leq\ell$).

\item The general matrix \G{j,k}\ is defined by
\[\G{j,k}=\diag\{\lt{j},\ldots,\lt{k}\}.\]
\E{j,k}, \F{j,k}\ and \HA{j,k}\ are defined to be variants of \G{j,k}. In the
case of \E{j,k}\ it is required that $\mu_{j},\ldots,\mu_{k}$ be even. In the
case
of \F{j,k}\ the requirement is that they be odd and in the case of \HA{j,k}\
that
they
be zero. Furthermore \Eh{j,k}, \Gh{j,k}\ and \Hh{j,k}\ are defined to be the
variants obtained by setting $\lt{j}=1$ in \E{j,k}, \G{j,k}\ and \HA{j,k}\
respectively.

\item The general matrices \LA{j,k}, \M{j,k}, and \N{j,k}\ are defined such
that
\[\begin{array}{l}
\{(L_{j,k})_{ab},(M_{j,k})_{ab},(N_{j,k})_{ab}\}=\\
\left\{\begin{array}{l}
\{\lt{a},\lt{a},\lt{a}\} \; \mbox{for }a=b-1;(a-j)\mbox{ even}\\
\{\lti{a-1},(-1)^{\mu_{a-1}}\lti{a-1},\lbd_{a-2}^{-1}t^{\mu_{a-1}}\} \;
\mbox{for }a=b+1;(a-j)\mbox{ even}\\
\{0,0,0\} \; \mbox{otherwise,}\end{array}\right.\end{array}\]
where $a,b$ have the index set $\{j,\ldots,k\}$.

\item The general matrices \Lp{j,k}, \Mp{j,k}, and \Np{j,k}\ are defined by
\[\begin{array}{l}
\{(L'_{j,k})_{ab},(M'_{j,k})_{ab},(N'_{j,k})_{ab}\}=\\
\left\{\begin{array}{l}
\{1,1,1\} \; \mbox{for }a=j;b=j+1\\
\{\lt{a},\lt{a},\lt{a}\} \; \mbox{for }a=b-1;a\neq j;(a-j)\mbox{ even}\\
\{\lt{2},\lt{2},\lbd_{2}\} \; \mbox{for }a=2;b=1\\
\{\lbd_{2}\lbd_{a-1}^{-1}t^{\mu_{2}-\mu_{a-1}},(-1)^{\mu_{a-1}}\lbd_{2}\lbd_{a-
1}^{-1}t^{\mu_{2}-\mu_{a-1}},
\lbd_{2}\lbd_{a-1}^{-1}t^{\mu_{a-1}}\} \;
a=b+1;a\neq j;(a-j)\mbox{ even}\\
\{0,0,0\} \; \mbox{otherwise,}\end{array}\right.\end{array}\]
where in \Mp{j,k}\ the additional constraint is imposed that $\mu_{2}$ be even.
The matrices \LL{j,k}\ are defined to be of the form \Lp{j,k}\ but such that
$\mu_{j}=\mu_{j+2}=\cdots=\mu_{k-2}=\mu_{k}=0;\mu_{2}=1$.

\item Let \WW{j,k}\ be the $(k-j+1)\times(k-j+1)$ matrix defined by
\[\WW{j,k}=\diag\{1,-1,\ldots,1,-1\} ,\]
where $k$ and $j$  are such that $(k-j)$ is even.

\item For the quantities $\mu_{j}$ (where $1\leq j\leq\ell$)
that are encountered later on,
quantities $\rho_{j}$ are defined to be functions of the $\mu_{j}$. Let
\[\rho_{j}=\left\{\begin{array}{ll}
					1&\mbox{if $\mu_{j}$ is odd}\\
					0&\mbox{if $\mu_{j}$ is
even.}\end{array}\right.\]
Furthermore, $\sigma_{j}$ is defined in terms of the above by letting
$\sigma_{j}=\arf(\rho_{j}-\mu_{j})$.

\end{enumerate}

In general, throughout this paper, otherwise undefined quantities such as
$\alpha,\lambda$ are taken to be non-zero complex numbers. In expressions
like $\alpha t^{\beta}$ therefore,  $\alpha$ is taken to be a non-zero
complex number, with $\beta$ being implicitly assumed to be an integer.

\end{subsection}
\end{section}
\begin{section}{The Weyl group of \Cl}
\label{sec-Weyl}

 The following list of the representatives of each conjugacy class of
involutions of
\W, where \W\
is the Weyl group of \Cl\, was obtained using the results of
Richardson(\cite{Ric}). As in the
case of \Bl\, the representatives lend themselves to listing by ``families'',
although
such a
listing is slightly more complicated for the present analysis. In this section
the
quantities $q$
and $r$ take integer values between 0 and $\ell$. For some families
$q$ and $r$ will have other restrictions placed upon them. Thus, by taking all
possible values of $q$
and $r$ (subject to restrictions), the conjugacy class representatives are
obtained.
In the
following list the numbers 1,2,$\ldots$,9 refer to the ``families''. \T\ is the
most
general form of
representative in each family. These results used in the next section.
\begin{enumerate}
\item This family contains just one representative given by
\[\begin{array}{llll}
\T(\A{j})&=&\A{j}&j=1,2,\ldots,\ell.
\end{array}\]
\item This family contains only the representative given by
\[\begin{array}{llll}
\T(\A{j})&=&-\A{j}&j=1,2,\ldots,\ell.
\end{array}\]
\item Let $q$ range over the integers 2,3,$\ldots,\ell-1$
and let \T\ be given by
\[\label{W.3}\begin{array}{llll}
\T(\A{j})&=&\A{j}&j<q-1\\
\T(\A{q-1})&=&\A{q-1}+2(\A{q}+\cdots+\A{\ell-1})+\A{\ell}& \\
\T(\A{k})&=&-\A{k}&k\geq q.
\end{array}\]
In the case where $q=2$, (\ref{W.3}) simplifies to
\[\begin{array}{llll}
\T(\A{1})&=&\A{1}+2(\A{2}+\cdots+\A{\ell-1})+\A{\ell}& \\
  \T(\A{k})&=&-\A{k}&k\geq q.
 \end{array}\]
\item Let $q$ and $r$ be such that $q$ is odd, $r\neq\ell$ and
$r-q>2$. \T\ is given by
\[\begin{array}{llll}
\T(\A{j})&=&-\A{j}&\mbox{$j$ is odd and }1\leq j\leq q\\
\T(\A{k})&=&\A{k-1}+\A{k}+\A{k+1}&\mbox{$k$ is even and }2\leq k\leq q-1\\
\T(\A{q+1})&=&\A{q}+\A{q+1}& \\
\T(\A{m})&=&\A{m}&q+2\leq m\leq r-2\\
\T(\A{r-1})&=&\A{r-1}+2(\A{r}+\cdots+\A{\ell-1})+\A{\ell}& \\
\T(\A{n})&=&-\A{n}& r\leq n\leq\ell.
\end{array}\]
\item Let $q$ be odd and such that $\ell-2>q$. Then
\[\begin{array}{llll}
\T(\A{j})&=&-\A{j}&\mbox{$j$ is odd and }1\leq j\leq q\\
\T(\A{k})&=&\A{k-1}+\A{k}+\A{k+1}&\mbox{$k$ is even and }1<k<q\\
\T(\A{q+1})&=&\A{q}+\A{q+1}+2(\A{q+2}+\cdots+\A{\ell-1})+\A{\ell}& \\
\T(\A{m})&=&-\A{m}&q+2\leq m\leq\ell.
\end{array}\]
\item Let $q$ be odd and such that $q\neq\ell$ if $\ell$ is odd and
also such that $q\neq\ell-1$ if $\ell$ is even. Then
\[\begin{array}{llll}
\T(\A{j})&=&-\A{j}&\mbox{$j$ is odd and }1\leq j\leq q\\
\T(\A{k})&=&\A{k-1}+\A{k}+\A{k+1}&\mbox{$k$ is even and }1<k<q\\
\T(\A{q+1})&=&\A{q}+\A{q+1}& \\
\T(\A{m})&=&\A{m}&q+2\leq m\leq\ell.
\end{array}\]
\item In this case $\ell-q$ is even, $q\neq 1$ and $q\neq 2$.
\[\begin{array}{llll}
\T(\A{j})&=&\A{j}&1\leq j\leq q-2\\
\T(\A{q-1})&=&\A{q-1}+A{q}& \\
\T(\A{k})&=&-\A{k}&q\leq l\leq\ell\mbox{ and }\ell-k\mbox{ is even}\\
\T(\A{m})&=&\A{m-1}+\A{m}+\A{m+1}&q<m<\ell\mbox{ and }\ell-m\mbox{ is
odd.}
\end{array}\]
\item This root transformation is associated only with odd values of $\ell$.
\[\begin{array}{llll}
\T(\A{j})&=&-\A{j}&1\leq j\leq\ell\mbox{ and $j$ is odd}\\
\T(\A{k})&=&\A{k-1}+\A{k}+\A{k+1}&1<k<\ell\mbox{ and $k$ is even.}
\end{array}\]
\item This root transformation is associated only with even values of $\ell$.
\[\begin{array}{llll}
\T(\A{j})&=&-\A{j}&1\leq j\leq\ell\mbox{ and $j$ is odd}\\
\T(\A{k})&=&\A{k-1}+\A{k}+\A{k+1}&1<k<\ell\mbox{ and $k$ is even}\\
\T(\A{\ell})&=&2\A{\ell-1}+\A{\ell}.&
\end{array}\]
\end{enumerate}
\end{section}
\begin{section}{Listing of involutive automorphisms of \Cli}
\label{sec-listing}
\begin{subsection} {Outline of method}

There are essentially two stages in determining the conjugacy classes of an
untwisted affine
Kac-Moody algebra ${\tilde{\cal{L}}}$ in the matrix formulation. The first is
to
find, for each type
of automorphism, all the \Ut\ matrices corresponding to Cartan preserving
involutive automorphisms.
This involves working through all the root preserving transformations of
${\tilde{\cal{L}}}^{0}$ in
the manner described in detail in \cite{Co2,Co3,Co4,Cla}. For \Cl\ these root
preserving
transformations are precisely those listed in the Section~\ref{sec-Weyl}. The
resulting \Ut\
matrices are listed in the following subsection. The second stage, which is
much
more difficult, is
to determine which of the corresponding automorphisms of ${\tilde{\cal{L}}}$
are
actually
conjugate to each other. Some general considerations on this matter for \Cli\
are
described in
the next subsection, and these are developed further for type 1a involutive
automorphisms with $u=1$ in Section~\ref{sec-1a1}, for type 1a involutive
automorphisms with $u=-1$ in Section~\ref{sec-1a2}, and for type 2a involutive
automorphisms (with $u=1$) in Section~\ref{sec-2a1}.
\end{subsection}

\begin{subsection} {Initial listing}
(i)  For the type 1a involutive automorphisms with $u=1$ the
\Ut\ matrices are:
\begin{equation}\label{A1}
\dsum\{\Dz{1,\ell},\Dz{1,\ell}\}\end{equation}
\begin{equation}\label{A2}\dsum\{\Dz{1,\ell},\Dz{1,\ell}\}\end{equation}
\begin{equation}\label{A3}
\offsum\{\Gh{1,\ell},\lt{\ell+1}\Ghi{1,\ell}\}
\end{equation}
\begin{equation}\label{A4}
\left(\begin{array}{cccc}
\Dz{1,q-1}&\zo&\zo&\zo\\
\zo&\zo&\zo&\G{q,\ell}\\
\zo&\zo&-\Dz{1,q-1}&\zo\\
\zo&\Gi{q,\ell}&\zo&\zo \end{array}\right)\end{equation}
\begin{equation}\label{A5}
\left(\begin{array}{cccccc}
\LA{1,q}&\zo&\zo&\zo&\zo&\zo\\
\zo&\Dz{q+2,r-1}&\zo&\zo&\zo&\zo\\
\zo&\zo&\zo&\zo&\zo&\G{r,\ell}\\
\zo&\zo&\zo&-\Lt{1,q}&\zo&\zo\\
\zo&\zo&\zo&\zo&-\Dz{q+2,r-1}&\zo\\
\zo&\zo&\Gi{r,\ell}&\zo&\zo&\zo
\end{array}\right)\end{equation}
\begin{equation}\label{A6}
\left(\begin{array}{cccc}
\Lp{1,q}&\zo&\zo&\zo\\
\zo&\zo&\zo&\G{q+2,\ell}\\
\zo&\zo&-\Ltp{1,q}&\zo\\
\zo&\lt{2}\Gi{q+2,\ell}&\zo&\zo\end{array}\right)\end{equation}
\begin{equation}\label{A7}
\dsum\{\LA{1,q},\Dz{q+2,\ell},\Lt{1,q},\Dz{q+2,\ell}\}\end{equation}
\begin{equation}\label{A8}\dsum\{\LA{1,q},\Dz{q+2,\ell},-\Lt{1,q},-
\Dz{q+2,\ell}\}
\end{equation}
\begin{equation}\label{A9}\left(\begin{array}{cccccc}
\Dz{1,q-1}&\zo&\zo&\zo&\zo&\zo\\
\zo&\LA{q,\ell-2}&\zo&\zo&\zo&\zo\\
\zo&\zo&\zo&\zo&\zo&\G{\ell,\ell}\\
\zo&\zo&\zo&-\Dz{1,q-1}&\zo&\zo\\
\zo&\zo&\zo&\zo&-\Lt{q,\ell-2}&\zo\\
\zo&\zo&\Gi{\ell,\ell}&\zo&\zo&\zo
\end{array}\right)\end{equation}
\begin{equation}\label{A10}
\left(\begin{array}{cccc}
\Lp{1,\ell-2}&\zo&\zo&\zo\\
\zo&\zo&\zo&\G{\ell,\ell}\\
\zo&\zo&-\Ltp{1,\ell-2}&\zo\\
\zo&\lt{2}\Gi{\ell,\ell}&\zo&\zo
\end{array}\right)\end{equation}
\begin{equation}\label{A11}
\dsum\{\Lp{1,\ell-1},\Ltp{1,\ell-1}\}\end{equation}
\begin{equation}\label{A12}
\dsum\{\Lt,-\Ltp{1,\ell-1}\}.\end{equation}

(ii)  For the type 1a involutive automorphisms with $u=-1$ the \Ut\ matrices
are:
\begin{equation}\label{B1}\dsum\{\Dz{1,\ell},\Dz{1,\ell}\}\end{equation}
\begin{equation}\label{B2}\dsum\{\Dz{1,\ell},-\Dz{1,\ell}\}\end{equation}
\begin{equation}\label{B3}
\left(\begin{array}{cc}
\zo&\Eh{1,\ell}\\ \lt{\ell+1}\Ehi{1,\ell}&\zo
\end{array}\right)\end{equation}
\begin{equation}\label{B4}
\left(\begin{array}{cccc}
\Dz{1,q-1}&\zo&\zo&\zo\\
\zo&\zo&\zo&\E{q,\ell}\\
\zo&\zo&-\Dz{1,q-1}&\zo\\
\zo&\Ei{q,\ell}&\zo&\zo \end{array}\right)\end{equation}
\begin{equation}\label{B5}\left(\begin{array}{cccc}
\Dz{1,q-1}&\zo&\zo&\zo\\
\zo&\zo&\zo&\F{q,\ell}\\
\zo&\zo&-\Dz{1,q-1}&\zo\\
\zo&\Fi{q,\ell}&\zo&\zo \end{array}\right)
\end{equation}
\begin{equation}\label{B6}
\left(\begin{array}{cccccc}
\M{1,q}&\zo&\zo&\zo&\zo&\zo\\
\zo&\Dz{q+2,r-1}&\zo&\zo&\zo&\zo\\
\zo&\zo&\zo&\zo&\zo&\E{r,\ell}\\
\zo&\zo&\zo&-\Mti{1,q}&\zo&\zo\\
\zo&\zo&\zo&\zo&-\Dz{q+2,r-1}&\zo\\
\zo&\zo&\Ei{r,\ell}&\zo&\zo&\zo
\end{array}\right)\end{equation}
\begin{equation}\label{B7}\left(\begin{array}{cccccc}
\M{1,q}&\zo&\zo&\zo&\zo&\zo\\
\zo&\Dz{q+2,r-1}&\zo&\zo&\zo&\zo\\
\zo&\zo&\zo&\zo&\zo&\F{r,\ell}\\
\zo&\zo&\zo&-\Mti{1,q}&\zo&\zo\\
\zo&\zo&\zo&\zo&-\Dz{q+2,r-1}&\zo\\
\zo&\zo&\Fi{r,\ell}&\zo&\zo&\zo
\end{array}\right)
\end{equation}
\begin{equation}\label{B8}
\left(\begin{array}{cccc}
\Mp{1,q}&\zo&\zo&\zo\\
\zo&\zo&\zo&\F{q+2,\ell}\\
\zo&\zo&-\Mtp{1,q}&\zo\\
\zo&\-\lt{2}\Fi{q+2,\ell}&\zo&\zo\end{array}\right)\end{equation}
\begin{equation}\label{B9}\left(\begin{array}{cccc}
\Mp{1,q}&\zo&\zo&\zo\\
\zo&\zo&\zo&\E{q+2,\ell}\\
\zo&\zo&-\Mtp{1,q}&\zo\\
\zo&\lt{2}\Ei{q+2,\ell}&\zo&\zo\end{array}\right)\end{equation}
\begin{equation}\label{B10}
\dsum\{\M{1,q},\Dz{q+2,\ell},\Mti{1,q},\Dz{q+2,\ell}\}\end{equation}
\begin{equation}\label{B11}\dsum\{\M{1,q},\Dz{q+2,\ell},-\Mti{1,q},-
\Dz{q+2,\ell}\}
\end{equation}
\begin{equation}\label{B12}
\left(\begin{array}{cccccc}
\Dz{1,q-1}&\zo&\zo&\zo&\zo&\zo\\
\zo&\M{q,\ell-2}&\zo&\zo&\zo&\zo\\
\zo&\zo&\zo&\zo&\zo&\F{\ell,\ell}\\
\zo&\zo&\zo&\Dz{1,q-1}&\zo&\zo\\
\zo&\zo&\zo&\zo&\Mti{q,\ell-2}&\zo\\
\zo&\zo&-\Fi{\ell,\ell}&\zo&\zo&\zo
\end{array}\right)\end{equation}
\begin{equation}\label{B13}\left(\begin{array}{cccccc}
\Dz{1,q-1}&\zo&\zo&\zo&\zo&\zo\\
\zo&\M{q,\ell-2}&\zo&\zo&\zo&\zo\\
\zo&\zo&\zo&\zo&\zo&\E{\ell,\ell}\\
\zo&\zo&\zo&-\Dz{1,q-1}&\zo&\zo\\
\zo&\zo&\zo&\zo&-\Mti{q,\ell-2}&\zo\\
\zo&\zo&-\Ei{\ell,\ell}&\zo&\zo&\zo
\end{array} \right)
\end{equation}
\begin{equation}\label{B14}
\left(\begin{array}{cccc}
\Mp{1,\ell-2}&\zo&\zo&\zo\\
\zo&\zo&\zo&\F{\ell,\ell}\\
\zo&\zo&\lt{2}\Mtpi{1,\ell-2}&\zo\\
\zo&-\lt{2}\Fi{\ell,\ell}&\zo&\zo
\end{array}\right)\end{equation}
\begin{equation}\label{B15}\left(\begin{array}{cccc}
\Mp{1,\ell-2}&\zo&\zo&\zo\\
\zo&\zo&\zo&\E{\ell,\ell}\\
\zo&\zo&-\lt{2}\Mtpi{1,\ell-2}&\zo\\
\zo&\lt{2}\Ei{\ell,\ell}&\zo&\zo
\end{array}\right)
\end{equation}
\begin{equation}\label{B16}
\dsum\{\Lp{1,\ell-1},\lbd_{2}t^{\mu_{2}}\Lpt{1,\ell-1}\}\end{equation}
\begin{equation}\label{B17}
\dsum\{\Lp{1,\ell-1},-\lbd_{2}t^{\mu_{2}}\Lpt{1,\ell-1}\}\end{equation}

(iii)  For the type 2a involutive automorphisms (with $u=1$) the \Ut\ matrices
are:
\begin{equation}\label{C1}
\dsum\{\D{1,\ell},\D{1,\ell}\}\end{equation}
\begin{equation}\label{C2}\dsum\{\D{1,\ell},\D{1,\ell}\}\end{equation}
\begin{equation}\label{C3}
\left(\begin{array}{cc}
\zo&\Hh{1,\ell}\\ \lt{\ell+1}\Hhi{1,\ell}&\zo\end{array}\right)\end{equation}
\begin{equation}\label{C4}
\left(\begin{array}{cccc}
\D{1,q-1}&\zo&\zo&\zo\\
\zo&\zo&\zo&t^{\mu_{q}}\HA{q,\ell}\\
\zo&\zo&-t^{2\mu_{q}}\Di{1,q-1}&\zo\\
\zo&t^{\mu_{q}}\Hi{q,\ell}&\zo&\zo \end{array}\right)\end{equation}
\begin{equation}\label{C5}
\left(\begin{array}{cccccc}
\N{1,q}&\zo&\zo&\zo&\zo&\zo\\
\zo&\D{q+2,r-1}&\zo&\zo&\zo&\zo\\
\zo&\zo&\zo&\zo&\zo&t^{\mu_{r}}\HA{r,\ell}\\
\zo&\zo&\zo&-t^{2\mu_{r}}\Nti{1,q}&\zo&\zo\\
\zo&\zo&\zo&\zo&-t^{2\mu_{r}}\Di{q+2,r-1}&\zo\\
\zo&\zo&t^{\mu_{r}}\Hi{r,\ell}&\zo&\zo&\zo
\end{array}\right)\end{equation}
\begin{equation}\label{C6}
\left(\begin{array}{cccc}
\Np{1,q}&\zo&\zo&\zo\\
\zo&\zo&\zo&t^{\mu_{\ell}}\HA{q+2,\ell}\\
\zo&\zo&-\lambda_{2}t^{2\mu_{\ell}}\Ntpi{1,q}&\zo\\
\zo&\-\lambda_{2}t^{\mu_{\el}}\Hi{q+2,\el}&\zo&\zo\end{array}\right)
\end{equation}
\begin{equation}\label{C7}
\dsum\{\N{1,q},\D{q+2,\ell},t^{\mu_{\ell+1}}\Nti{1,q},t^{\mu_{\ell+1}}\D{q+2,\
ell}
\end{equation}
\begin{equation}\label{C8}
\left(\begin{array}{cccccc}
\D{1,q-1}&\zo&\zo&\zo&\zo&\zo\\
\zo&\N{q,\ell-2}&\zo&\zo&\zo&\zo\\
\zo&\zo&\zo&\zo&\zo&\G{\ell,\ell}\\
\zo&\zo&\zo&-t^{2\mu_{\ell}}\Di{1,q-1}&\zo&\zo\\
\zo&\zo&\zo&\zo&-t^{2\mu_{\ell}}\Nti{q,\ell-2}&\zo\\
\zo&\zo&t^{2\mu_{\ell}}\Gi{\ell,\ell}&\zo&\zo&\zo
\end{array}\right)\end{equation}
\begin{equation}\label{C9}
\left(\begin{array}{cccc}
\Np{1,\ell-2}&\zo&\zo&\zo\\
\zo&\zo&\zo&\G{\ell,\ell}\\
\zo&\zo&-\lambda_{2}^{\mu_{\ell}}\Ntpi{1,\ell-2}&\zo\\
\zo&\lambda_{2}^{\mu_{\ell}}\G{\ell,\ell}&\zo&\zo
\end{array}\right)\end{equation}
\begin{equation}\label{C10}
\dsum\{\Np{1,\ell-1},\Np{1,\ell-1}\}\end{equation}
\begin{equation}\label{C11}
\dsum\{\Np{1,\ell-1},-\Np{1,\ell-1}\}.\end{equation}
\end{subsection}

\begin{subsection}{Simplification of listing}
In this subsection, we will investigate the conjugacy of some of the
automorphisms specified in the previous subsection.

(i) Let \Ut\ given by (\ref{A5}) and let \Upt\
be obtained from \Ut\ by replacing \LA{1,q}\ with \WW{1,q}. We define
\[\St=\dsum\{\V{1}(t),\ldots,\V{\frac{1}{2}(q+1)},
\I{\el-q-1},\Vp{1},\ldots,\Vp{\frac{1}{2}(q+1)},\I{\el-q-1}\},\]
where
\[\V{j}=\frac{1}{\sqrt{2}}\left(\begin{array}{cc}
\lt{2j-1}&\lt{2j-1}\\1&-1\end{array}\right),
\Vp{j}=\frac{1}{\sqrt{2}}\left(\begin{array}{cc}\lti{2j-1}&\lti{2j-1}\\1&-
1\end{array}\right).\]
Thus
\[ \St\Upt\Sti=\Ut \; , \; \tilde{{\bf S}}(t)\J\St=\J. \]
Consequently in (\ref{A5}) we may assume without loss of generality that
\LA{1,q}\ may
be replaced by \WW{1,q}. Similarly in (\ref{A7}), (\ref{A8}) and (\ref{A9}) we
may assume that the submatrices \LA{j,k}\ are replaced by the submatrices
\WW{j,k}. Furthermore, the above analysis may be modified slightly for type 1a
automorphisms with $u=-1$ and also for type 2a automorphisms with $u=1$.
Thus, if
\Ut\ is given by any one of (\ref{B6}), (\ref{B7}), (\ref{B10}), (\ref{B11}),
(\ref{B12}),
 (\ref{C5}), (\ref{C7}) or (\ref{C8}), then the submatrices \M{j,k}\ (or
\N{j,k})
may be replaced by \WW{j,k}.

(ii) Let \Ut\ be given by \bref{A3}\ and let \Upt\ be obtained from \Ut\ by
setting
$\lambda_{j}=1$ (for $2\leq j\leq\ell+1$) and also by setting
$\mu_{j}=\rho_{j}$
 (for $2\leq j\leq \ell+1$). If we define
\[{\bf t}=\lambda_{\ell+1}^{-\frac{1}{4}}\diag\{1,\lambda_{2}^{\frac{1}{2}}
t^{\sigma_{2}},\ldots,\lambda_{\ell}^{\frac{1}{2}}t^{\sigma_{\ell}}\} \; ,
\;\St=\dsum\{
t^{\sigma_{\ell+1}}{\bf t},{\bf t}^{-1}\},\]
then
\[\tilde{{\bf S}}(t)\J\St=t^{\sigma_{\ell+1}}\J \; , \;
\St\Upt\Sti=\lambda_{\ell+1}^{-\frac{1}{2}}t^{\sigma_{\ell+1}}\Ut.\]
Thus in \bref{A3}\ we may assume that $\lambda_{j}t^{\mu_{j}}=t^{\rho_{j}}$
(for
 $2\leq j\leq\ell+1$). Furthermore the above argument may be modified slightly
for the matrices given by \bref{B3}\ or \bref{C3}\ but with the same result.
In fact this result may be extended even further. If \Ut\ is given by any one
of \bref{A4}, \bref{A5}, \bref{A9}, \bref{B4}, \bref{B5}, \bref{B6}, \bref{B7},
\bref{B12}, \bref{B13}, \bref{C4}, \bref{C5}, \bref{C6}, \bref{C8}\ or
\bref{C9},
then in the submatrices \E{j,k}\ (or \F{j,k},\G{j,k},\HA{j,k}\ etc.) we may
assume that \mbox{$\lt{m}=t^{\rho_{m}}$} for $j\leq m\leq k$.

(iii) Let \Ut\ be given by \bref{A6}\ and let \Upt\ be obtained from \Ut\ by
putting $\lt{j}=1$ (for $3\leq j\leq q,(q-j)$ is even) and by putting
$\lt{j}=t^{\rho_{j}}$
 (for $j=2,q+2\leq j\leq\ell$). Then let
\[{\bf t}_{1}=\diag\{\lambda_{2}^{-\frac{1}{2}}t^{-\sigma_{2}},1,\lambda_{2}^{-
\frac{1}{2}}
\lambda_{3}t^{\mu_{3}-\sigma_{2}},1,\ldots,\lambda_{2}^{-
\frac{1}{2}}\lambda_{q}
t^{\mu_{q}-\sigma_{2}},1\},\]
\[{\bf t}_{2}=\lambda_{2}^{-
\frac{1}{4}}\diag\{\lambda_{q+2}^{\frac{1}{2}}t^{\sigma_{q+2}},
\ldots,\lbd_{\ell}^{\frac{1}{2}}t^{\sigma_{\ell}}\},\]
\[{\bf
t}_{3}=\diag\{\lbd_{2}^{\frac{1}{2}},t^{\sigma_{2}},\lbd_{2}^{\frac{1}{2}}\lbd_{3}^{
-1}
t^{-\mu_{3}},t^{\sigma_{2}},\ldots,\lbd_{2}^{\frac{1}{2}}\lbd_{q}^{-1}t^{-\mu_{q}},
t^{\sigma_{2}}\},\]
and
\[\St=\dsum\{{\bf t}_{1},{\bf t}_{2},{\bf t}_{3},{\bf t}_{2}^{-1}\}.\]
Then $\StJSt=t^{\sigma_{2}}\J$, and $\St\Upt\Sti=t^{\sigma_{2}}\Ut$. Hence,
in
\bref{A6}, we may assume that the matrices \Ut\ are all of the form \Upt,
where \Upt\ is as defined above. Similar assumptions may be made about the
matrices given by \bref{A10}, \bref{B8}, \bref{B9}, \bref{B14}, \bref{B15},
\bref{C6}\ and
\bref {C9}. The proof is very similar to that just given.

(iv) Let \Ut\ be any matrix of the form $\dsum\{{\bf a},{\bf b}\}$, where
{\bf a} and {\bf b} are $(\ell\times\ell)$ matrices such that
\[{\bf a}=\diag\{a_{1},\ldots,a_{\ell}\} \; ,\;{\bf
b}=\diag\{b_{1},\ldots,b_{\ell}\},\]
and let ${\bf a}'$ and ${\bf b}'$ be obtained from {\bf a} and {\bf b} by
exchanging
$j$ and $k$ (where $1\leq j,k\leq\ell$) in their respective index sets.
Let \Upt\ be given by \mbox{\offsum$\{\m{a}',\m{b}'\}$} and define {\bf S} by
\[S_{ab}=\left\{\begin{array}{ll}
1&\mbox{if }a=b,a\neq j,k,j+\ell,k+\ell\\
0&\mbox{if }a=b,a=j,k,j+\ell,k+\ell\\
1&\mbox{if }(a,b)=(j,k),(k,j),(j+\ell,k+\ell),(k+\ell,j+\ell)\\
0& \mbox{otherwise}.\end{array}\right.\]
Then $\SJS=\J$ and ${\bf S}\Ut{\bf S}^{-1}=\Upt.$
Similarly if
\[\Ut=\offsum\{\m{a},\m{b}\}\]
\[\Upt=\offsum\{\m{a}',\m{b}'\},\]
then we still find that
\[\SJS=\J \; , \;{\bf S}\Ut{\bf S}^{-1}=\Upt.\]
This may be extended further to some of the other matrices under consideration.
In fact, \Ut\ will almost always have submatrices that are diagonal. The
above analysis will extend to these cases, allowing us to alter the index sets
of
the diagonal parts arbitrarily.

(v) Let $\Ut=\dsum\{{\bf a},{\bf b}\}$, where {\bf a} and {\bf b} are as
defined in (iv), and let \Upt\ be obtained from \Ut\ by exchanging the
elements $a_{j}$ and $b_{j}$ for $1\leq j\leq\ell$. Define {\bf S}
by
\[S_{ab}=\left\{\begin{array}{ll}
1&\mbox{if }a=b,a\neq j,j+\ell\\
0&\mbox{if }a=b=j,j+\ell\\
i&\mbox{if }(a,b)=(j,j+\ell),(j+\ell,j)\\
0& \mbox{otherwise}.\end{array}\right.\]
Then $\SJS=\J$ and ${\bf S}\Ut{\bf S}^{-1}=\Upt.$ Thus the order of diagonal
elements may be regarded as being arbitrary, up to a certain point.
Furthermore,
if \Ut\ is given by
\[\Ut=\offsum\{\m{a},\m{b}\},\]
and if \Upt\ is obtained from \Ut\ by exchanging the elements $a_{j}$ and
$b_{j}$, then the above analysis still holds. That is, $a_{j}$ and $b_{j}$
may be exchanged without loss of generality. As in (iv), we may extend this to
those
matrices \Ut\ with diagonal submatrices. The conclusion still holds. That is,
in the index set $\{1,\ldots,2\ell\}$ of \Ut\ we may exchange the $j$th and the
$(j+\ell)$th members, without loss of generality.

(vi) Let \Ut\ be of the general form \bref{C4}\ and let \Upt\ be obtained form
\Ut\
by setting $\mu_{q}=0$. If we let \mbox{$\St=\dsum\{\I{\ell},t^{-\mu_{q}}
\I{\ell}\}$}, then
\[ \St\Ut\Stii=\Upt \; ,\; \SJS=t^{-\mu_{q}}. \]
Thus we may assume without loss of generality that $\mu_{q}=0$. Similarly,
if \Ut\ is given by \bref{C5}, \bref{C6}, \bref{C8}\ or \bref{C9}\, then
without
loss of
generality the quantities $\mu_{r}$, $\mu_{\ell}$, and $\mu_{\ell}$ may be
assumed to be zero.

(vii) Let \Ut\ be of the form
\[\Ut=\left(\begin{array}{cccc}
\m{H}_{1}&\zo&\zo&\zo\\
\zo&\zo&\zo&{\bf a}\\
\zo&\zo&\m{H}_{2}&\zo\\
\zo&{\bf a}^{-1}&\zo&\zo\end{array}\right),\]
where $\m{H}_{1}$ and $\m{H}_{2}$ are arbitrary $(\ell-q)\times(\ell-q)$
matrices and
\[{\bf a}=\diag\{t^{\mu_{\ell-q+1}},\ldots,t^{\mu_{\ell}}\}.\]
Let \Upt\ be given by \mbox{\dsum$\{\m{H}_{1},\I{q},\m{H}_{2},-\I{q}\}$}.
If we define \St\ by
\[\St=\left(\begin{array}{cccc}
\I{\ell-q}&\zo&\zo&\zo\\
\zo&(2)^{-\frac{1}{2}}\I{q}&\zo&-(2)^{-\frac{1}{2}}{\bf a}\\
\zo&\zo&\I{\ell-q}&\zo\\
\zo&(2)^{-\frac{1}{2}}{\bf
a}^{-1}&\zo&(2)^{-\frac{1}{2}}\I{q}\end{array}\right),\]
then we have that $\StJSt=\J$ and also that $\St\Upt\Sti=\Ut$. This allows
for a great amount of simplification later on. Many automorphisms can be
shown to be conjugate by using this matrix transformation. Moreover, it
is also valid for the type 1a involutive automorphisms (with $u=-1$) and
the type 2a involutive automorphisms (with $u=1$). This is because in the
former
case $\Smt=\St$ and in the latter $\m{S}(t^{-1})=\St$.

\end{subsection}
\end{section}

\begin{section}{Study of the type 1a involutive automorphisms with $u=1$}
\label{sec-1a1}

\begin{subsection}{Conjugacy analysis}
Using the results of Section~\ref{sec-listing}, we find that each involutive
automorphism
of type 1a with $u=1$ is conjugate to at least one automorphism corresponding
to one of the following matrices:
\begin{equation}\label{AA1}
\Ut=\dsum\{\Dz{1,\ell},\Dz{1,\ell}\}\end{equation}
\begin{equation}\label{AA2}
\Ut=\dsum\{\I{\ell},-\I{\ell}\}\end{equation}
\begin{equation}\label{AA3}
\Ut=\offsum\{\I{\ell},t\I{\ell}\}
\end{equation}
\begin{equation}\label{AA4}
\Ut=\left(\begin{array}{cccc}
\LL{1,q}&\zo&\zo&\zo\\
\zo&\zo&\zo&\I{\ell-q-1}\\
\zo&\zo&-\LLt{1,q}&\zo\\
\zo&t\I{\ell-q-1}&\zo&\zo\end{array}\right)
\end{equation}
\begin{equation}\label{AA5}
\Ut=\dsum\{\LL{1,\ell-1},\LLt{1,\ell-1}\}\end{equation}
\begin{equation}\label{AA6}
\Ut=\dsum\{\LL{1,\ell-1},-\LLt{1,\ell-1}\}.\end{equation}

We note first that the mapping $t\mapsto st$ is of some importance since
the conjugacy conditions for two type 1a involutive automorphisms (with $u=1$)
include expressions $\m{U}(st)$. However, we may always assume that $s=1$.
This is because the mapping $t\mapsto st$ corresponds merely to a change in
some of the parameters $\lbd_{j}$ which we have already shown to be
completely arbitrary (changing them does not affect the conjugacy class
of the associated automorphism). In particular, for those automorphisms
corresponding to
\bref{AA1}, it follows from (iv) of Section~\ref{sec-listing}
that the matrices \Ut\ may be assumed to be of the form
\[\Ut=\dsum\{\I{m},-\I{\ell-m},\I{m},-\I{\ell-m}\},\]
and we may assume that $\ell-m\leq [\frac{\ell}{2} ]$. Thus we are
investigating
precisely
$(1+[\frac{\ell}{2}])$ distinct automorphisms, each of which corresponds to a
different value
of $(\ell-m)$.

We define \iA{j}\ (for $0\leq j\leq[\frac{\ell}{2}]$) to be the conjugacy
class containing the type 1a (with $u=1$) automorphism corresponding to
${\bf U}_{j}$, where
\begin{equation}\label{E1a}
{\bf U}_{j}=\dsum\{\I{\ell-j},-\I{j},\I{\ell-j},-\I{j}\}.
\end{equation}
We will demonstrate that \iA{j}\ is disjoint from \iA{k}\ when $j\neq k$. If
the classes \iA{j}\ and \iA{k}\ did coincide for different values of $j$ and
$k$,
then it would be
necessary for a matrix \St\ to exist such that
\begin{equation}\label{E1b}
\St\;\dsum{\bf U}_{j}\;\Sti=\lambda t^{\mu}{\bf U}_{k},
\end{equation}
where ${\bf U}_{j}$ and ${\bf U}_{k}$ are defined as in \bref{E1a}. This would
have to hold
for all values of $t$, and for $t=1$ in particular. Upon putting $t=1$ in the
equation
\bref{E1b}\, we find that there cannot be any Laurent matrix \St\ that
satisfies
\bref{E1b}. Thus the automorphisms that correspond to \bref{A1}\ fall into
precisely $(1+[\frac{\ell}{2}])$ disjoint conjugacy classes which we have
called $\iA{0},\ldots,\iA{(1+\elbt)}.$ It is clear that \iA{0}\ contains only
the identity automorphism.

We move  onto the automorphism corresponding to \bref{AA2} and define
\iB\ to be the conjugacy class containing the automorphism
\mbox{$\Ut\,1,0\}$}, where \Ut\ is given by \bref{AA2}. We will show that \iB\
is disjoint from the classes
\iA{j}\ that we have discussed previously. If we suppose that \iB\ and \iA{j}\
are the same for some suitable $j$, then the following must hold
\[\St{\bf U}\Sti=\lambda t^{\mu}{\bf U}_{j},\]
where ${\bf U}_{j}$ is as defined in \bref{E1a}. If we substitute $t=1$ in this
equation, then it is easily seen that \el\ must be even. Furthermore when \el\
is
even, the matrix {\bf S}(1) cannot be chosen such that
\mbox{$\tilde{{\bf S}}(1)\J{\bf S}(1)=\lambda\J$}. Thus \iB\ is a class
that is disjoint from all of the others identified so far.

Next we consider the automorphism corresponding to \bref{AA3}.
Consideration of the determinant of \Ut\ ,where \Ut\ is given by \bref{AA3},
allows us to infer that the automorphism
corresponding to it does not belong to any of the conjugacy classes that we
have
identified
so far. Otherwise we would have
\begin{equation}\label{E4}
\St{\bf U}\Sti=\lbd t^{\mu}\Ut,
\end{equation}
where {\bf U} is a $t$-independent matrix. Taking determinants of both sides of
\bref{E4}\ implies that $k=-\arf$, which is obviously a contradiction. Hence
the
automorphisms corresponding to \bref{A3},
and for which $\mu_{\ell+1}$ is odd, belong to some previously unidentified
conjugacy class which we call \iC.

Let \Ut\ be given by \bref{AA4} and recall that the type 1a involutive
automorphism corresponding to
\mbox{$\offsum\{\I{\ell},t\I{\ell}\}$} belongs to the conjugacy class \iC.
Then we define a matrix \St\ by
\[\St=\left(\begin{array}{cccc}
{\bf Q}_{1}&\zo&{\bf Q}_{2}&\zo\\
\zo&\I{\ell-q-1}&\zo&\zo\\
{\bf Q}_{3}&\zo&{\bf Q}_{4}&\zo\\
\zo&\zo&\zo&\I{\ell-q-1}\end{array}\right),\]
where the $(ab)$th elements of the $(q+1)\times(q+1)$ submatrices are given by
\[\{({\bf Q}_{1}),({\bf Q}_{2}),({\bf Q}_{3}),({\bf Q}_{4})\}_{ab}=\left\{
\begin{array}{cc}
\{1,(2)^{-\frac{1}{2}},(2)^{-\frac{1}{2}},1\}&\mbox{for }a=b; a\mbox{ odd,}\\
\{-(2)^{-\frac{1}{2}}it,i,i,-(2)^{-\frac{1}{2}}it^{-1}\}&\mbox{for }a=b;
a\mbox{
even,}\\
\{i,-(2)^{-\frac{1}{2}}i,(2)^{-\frac{1}{2}}i,-i\}&\mbox{for }a=b-1; a\mbox{
odd,}\\
\{(2)^{-\frac{1}{2}}t,1,-1,-(2)^{-\frac{1}{2}}t^{-1}\}&\mbox{for }a=b-1;a\mbox{
even,}\\
\{0,0,0,0\}&\mbox{otherwise.}\end{array}\right.\]
Then $\StJSt=\J$ and $\St\;\offsum\{\I{\ell},t\I{\ell}\}\;\Sti=\Ut$. Hence the
automorphisms
\bref{AA4}\ all belong to the conjugacy class \iC.

The only remaining type 1a involutive automorphisms with $u=1$ are those given
by
\bref{AA5} and \bref{AA6}. Let us first examine those given by \bref{AA5}.
Consideration of the
determinant of \Ut\ of \bref{AA5} shows that the automorphism corresponding to
it cannot belong to
any of the conjugacy classes \iA{0},
$\ldots$,\iA{\elbt}\ or \iB. Either it belongs to \iC, or to some as yet
unidentified
conjugacy
class. Let us suppose that it belongs to the class \iC, so that for some matrix
\St
\begin{equation}\label{E5a}
\St\;\dsum\{\LL{1,\ell-1},\LLt{1,\ell-1}\}\;\Sti=\offsum\{\I{\ell},t\I{\ell}\}.
\end{equation}
We are assuming that \bref{E5a}\ holds for every non-zero value of $t$, and
in particular for $t=1$. Let {\bf T} be defined to be obtained from
\LL{1,\ell-1}\ by setting $t=1$. We know that there exists a
$t$-independent matrix {\bf R} such that
\[{\bf R}\;\dsum\{{\bf T},\tilde{{\bf T}}\}\;{\bf R}^{-1}=\dsum\{\WW{1,\ell-1}
,\WW{1,\ell-1}\},\]
but we know that there exists no $t$-independent matrix {\bf Q} such that {\em
both}
\[{\bf Q}\;\dsum\{\WW{1,\ell-1},\WW{1,\ell-1}\}\;{\bf Q}^{-
1}=\beta\offsum\{\I{\ell}
,\I{\ell}\}\]
{\em and}
\[ \tilde{{\bf Q}}\J{\bf Q}=\gamma\J.\]
Thus our assumption that it belongs to the class \iC\ is false, and so the
automorphism \bref{AA5}
does belong to some new conjugacy class \iD.

Finally, we examine the automorphism \bref{AA6} and
we demonstrate that this automorphism belongs to the conjugacy class
\iC. Let \Ut\ be given by \bref{AA6}\ and let \Upt\ be given by \bref{AA3}.
Then
\[\St\Upt\Sti=\Ut \; , \;\StJSt=\J,\]
where
\[\St=\left(\begin{array}{cc}
					{\bf Q}_{1}&{\bf Q}_{2}\\
						{\bf Q}_{3}&{\bf
Q}_{4}\end{array}\right).\]
The submatrices ${\bf Q}_{1},{\bf Q}_{2},{\bf Q}_{3}$ and ${\bf Q}_{4}$ are
of dimension $\ell\times\ell$ and their $(ab)$th elements are given by
\[\{({\bf Q}_{1}),({\bf Q}_{2}),({\bf Q}_{3}),({\bf Q}_{4})\}_{ab}=
\left\{\begin{array}{cc}
\{1,(2)^{-1/2},(2)^{-1/2},1\}&\mbox{for }a=b; a\mbox{ odd,}\\
\{-(2)^{-1/2}it,i,i,-(2)^{-1/2}it^{-1}\}&\mbox{for }a=b; a\mbox{ even,}\\
\{i,-(2)^{-1/2}i,(2)^{-1/2}i,-i\}&\mbox{for }a=b-1; a\mbox{ odd,}\\
\{(2)^{-1/2}t,1,-1,(2)^{-1/2}t^{-1}\}&\mbox{for }a=b+1; a\mbox{ even,}\\
\{0,0,0,0\}&\mbox{otherwise.}\end{array}\right.\]

\end{subsection}

\begin{subsection} {Explicit forms for the automorphisms}

With the analysis of the previous subsection we have identified all the
conjugacy
classes of the type 1a involutive
automorphisms within the group of all automorphisms of \Cli. Curiously, there
are more conjugacy classes for the case $\ell=2m$ than there are for the case
$\ell=2m+1$. This is due to the nature of the conjugacy classes of the Weyl
group
for the
algebra \Cl. We recall that for even $\ell$ there was one extra family of
conjugacy
classes of involutive Weyl group elements.
Some of the automorphisms associated with these Weyl group elements give rise
to
an `extra' conjugacy class of involutive automorphisms of \Cli.

\begin{enumerate}
\item The conjugacy class \iA{0}\ is clearly the conjugacy class which contains
only the identity automorphism.

\item The conjugacy class \iA{b}\ for $b\neq 0$ has a representative $\psi$,
where $\psi$ has the following effect upon the basis elements:
\begin{equation}\begin{array}{lll}
\psi(\hk{k})&=&\hk{k} \; \mbox{for }k=1,2,\ldots,\ell\\
\psi(e_{j\delta\pm\alpha_{H}})&=&e_{j\delta\pm\alpha_{H}}\\
\psi(e_{j\delta\pm\alpha_{k}})&=&\left\{\begin{array}{l}
					e_{j\delta\pm\alpha_{k}} \; \mbox{for
}k\neq\ell-b;1\leq k\leq\ell\\
					-e_{j\delta\pm\alpha_{k}} \; \mbox{for
}k=\ell-b\end{array}\right. \\
\psi(c)&=&c\\
\psi(d)&=&d .\end{array}\label{eq:Arep}\end{equation}
Its $\Ut$ matrix is given by the ${\bf U}_{j}$ of \bref{E1a}.

\item The conjugacy class \iB\ has the following representative, which
corresponds to
the matrix $\Ut=\mbox{\dsum$\{\I{\ell},-\I{\ell}\}$}$:
\begin{equation}\begin{array}{lll}
\psi(\hk{k})&=&\hk{k} \; \mbox{for }k=1,2,\ldots,\ell\\
\psi(e_{j\delta\pm\alpha_{H}})&=&-e_{j\delta\pm\alpha_{H}}\\
\psi(e_{j\delta\pm\alpha_{k}})&=&\left\{\begin{array}{l}
					e_{j\delta\pm\alpha_{k}} \; \mbox{for
}k=1,2,\ldots,\ell-1\\
					-e_{j\delta\pm\alpha_{\ell}} \; \mbox{for
}k=\ell\end{array}\right.\\
\psi(c)&=&c\\\psi(d)&=&d.\end{array}\label{eq:Brep}\end{equation}

\item The conjugacy class \iC\ has the following representative, which
corresponds to the matrix $\Ut=\mbox{\offsum$\{\I{\ell},t\I{\ell}\}$}$:
\begin{equation}\begin{array}{lll}
\psi(\hk{k})&=&\left\{\begin{array}{l}
		\hk{k} \; \mbox{for }k=1,2,\ldots,\ell-1\\
		\hk{\ell}+c \; \mbox{for }k=\ell\end{array}\right.\\
\psi(e_{j\delta\pm\alpha_{H}})&=&-e_{(j\pm 1)\delta\mp\alpha_{H}}\\
\psi(e_{j\delta\pm\alpha_{k}})&=&\left\{\begin{array}{l}
					e_{j\delta\mp\alpha_{k}} \; \mbox{for
}k=1,2,\ldots,\ell-1\\
					-e_{(j\pm 1)\delta\mp\alpha_{\ell}} \;
\mbox{for }k=\ell\end{array}\right.\\
\psi(c)&=&c\\
\psi(d)&=&d+4(\ell+1)(\sum_{p=1}^{\ell-1}p\hk{p})+\frac{\ell}{2}\hk{\ell}-
\frac{\ell(\ell+1)}{2}c.
\end{array}\label{eq:Crep}\end{equation}

\item The conjugacy class \iD\, which occurs only for even values of $\ell$,
has
the following representative, which is the automorphism corresponding
to the $\Ut$ matrix of \bref{AA5}:
\begin{equation}\begin{array}{lll}
\psi(\hk{k})&=&\hk{k}+(-1)^{k+1}c \; \mbox{for }k=1,2,\ldots,\ell\\
\psi(e_{j\delta\pm\alpha_{H}})&=&e_{(j+1)\delta\pm\alpha_{H}\mp
2\alpha{1}}\\
\psi(e_{j\delta\pm\alpha_{k}})&=&\left\{\begin{array}{l}
			-e_{(j\pm 1)\delta\mp\alpha_{k}} \; \mbox{for
}k=1,\ldots,\ell-1;k\mbox{ is odd}\\
			e_{(j\mp 1)\delta\pm\alpha_{k-
1}\pm\alpha_{k}\pm\alpha_{k+1}} \; \mbox{for }k=1,\ldots,\ell-2;k\mbox{ is
even}\\
e_{(j\mp 1)\delta\pm 2\alpha_{\ell-1}\pm\alpha_{\ell}} \; \mbox{for
}k=1\end{array}\right.\\
\psi(c)&=&c\\
\psi(d)&=&d+2(\ell+1)(\sum_{p=1,p\
odd}^{\ell-1}\hk{p})-\frac{\ell(\ell+1)}{2}c.
\end{array}\label{eq:Drep}\end{equation}

\end{enumerate}
\end{subsection}
\end{section}

\begin{section}{Study of the type 1a involutive automorphisms with $u=-1$}
\label{sec-1a2}

\begin{subsection}{Conjugacy analysis}
We shall begin this section by examining the automorphisms \bref{B1}.
Define
$s_{j}$ (for $2\leq j\leq\ell$) by letting $s_{j}=1$ if $\lbd_{j}=-1$, and by
letting
$s_{j}=0$ of
$\lbd_{j}=1$. Then, with
\[ \St
=\mbox{diag$\{1,t^{s_{2}},\ldots,t^{s_{\ell}},1,t^{-s_{2}},\ldots,t^{-s_{\ell}}\}$} ,\]
it follows that
\[ \StJSt=\J \; , \; \St\I{2\ell}\Smti=\Ut , \]
 so all of the
automorphisms \bref{B1}\ belong to the same conjugacy class, which we will call
\iE.

The automorphisms \bref{B2}\ all belong to the class \iE. If \Ut\ is given
by \bref{B2}\ then we may define \St\ by
\[\St=\diag\{1,t^{\rho_{2}},\ldots,t^{\rho_{\ell}},t,t^{1-\rho_{2}},\ldots,t^{1-
\rho_{\ell}}\},\]
so that
\[\StJSt=\J \; , \;\St\I{2\ell}\Smti=\Ut.\]

The analysis of Section~\ref{sec-listing} implies that that the automorphisms
\bref{B3}\ are
all conjugate and, furthermore, they all belong to the conjugacy class \iE.

Consider the type 1a automorphism with $s=1$ that corresponds to the matrix
\St\, where
\begin{equation}\label{F1}
\St=\dsum\{\I{\ell},t\I{\ell}\}.
\end{equation}
This automorphism is such that conjugation of an automorphism given by
\bref{B4}\ gives rise to an automorphism given by \bref{B5}. Hence every
automorphism given by \bref{B5}\ is conjugate to one given by \bref{B4}.
We infer from the results of the Section~\ref{sec-listing} that the
automorphisms
\bref{B4}\ belong to the conjugacy class \iE. Similar results follow for
the remaining type 1a involutive automorphisms (with $u=-1$). That is,
for those given by \bref{B6}\ through to \bref{B17}. The results of
Section~\ref{sec-listing}, together with conjugation by the automorphism
corresponding to
\bref{F1}\, imply that all of these automorphisms belong to the conjugacy class
\iE.

\end{subsection}

\begin{subsection}{Explicit form of automorphism}
A representative automorphism $\psi$ of the class \iE\ is given by
\begin{equation}
\begin{array}{lll}
\psi(\hk{k})&=&\hk{k} \; \mbox{for }k=1,\ldots,\ell\\
\psi(e_{j\delta\pm\alpha_{H}})&=&(-1)^{j}e_{j\delta\pm\alpha_{H}}\\
\psi(e_{j\delta\pm\alpha_{k}})&=&(-1)^{j}e_{j\delta\pm\alpha_{k}} \; \mbox{for
}k=1,\ldots,\ell\\
\psi(c)&=&c\\
\psi(d)&=&d.
\end{array}
\label{eq:Erep}\end{equation}
This automorphism corresponds to the matrix $\Ut = \I{2\ell}$.

\end{subsection}
\end{section}

\begin{section}{Study of the type 2a involutive automorphisms with $u=1$}
\label{sec-2a1}

\begin{subsection}{Conjugacy analysis}
Using the results of Section~\ref{sec-listing}, we find that each type 2a
involutive
automorphism (with $u=1$) is conjugate to some type 2a involutive automorphism
corresponding to one of:
\begin{equation}\label{CC1}
\Ut=\dsum\{\D{1,\ell},\Di{1,\ell}\}\end{equation}\begin{equation}\label{CC2}
\Ut=\dsum\{\D{1,\ell},t\Di{1,\ell}\}\end{equation}\begin{equation}\label{CC3}
\Ut=\dsum\{\D{1,\ell},-\Di{1,\ell}\}\end{equation}\begin{equation}\label{CC4}
\Ut=\dsum\{\D{1,\ell},-t\Di{1,\ell}\}\end{equation}

The conjugacy conditions (\ref{eq:B.202}) and (\ref{eq:B.208}) for type 2a
automorphisms may be
further refined in the involutive case as follows. Let two involutive
automorphisms
of type 2a and
with $u=1$ have associated matrices \Ut\ and \Upt. Then these are conjugate via
the automorphism
\mbox{$\{\St,s,\xi\}$} (where $s^{2}=1$)  if and only if
\begin{equation}\begin{array}{ll}\label{H1}
\St\m{U}(st)\Stii=\lbd t^{\mu}\Upt&\mbox{(where \St\ is of type 1a)}\\
\St\m{U}(st^{-1})\Stii=\lbd t^{\mu}&\mbox{(where \St\ is of type 2a).}
\end{array}\end{equation} Thus, to prove that any two automorphisms are not
conjugate, we may prove
that it is not possible to satisfy \bref{H1}. In addition, it may be possible
to satisfy
\bref{H1}\
but only if certain forbidden or contradictory conditions are fulfilled. In
this
subsection we prove
that automorphisms are not conjugate by substituting the values $s=1,-1$ and
$t=1,-1$ into an equation of the form \bref{H1}. We then find a number of
contradictions which
complete the proof. In some cases it may be that there is no matrix $\m{S}(1)$
or
$\m{S}(-1)$ that
satisfies \bref{H1}\ (with substitutions) and also satisfies
\[\tilde{\m{S}}(\pm 1)\J\m{S}(\pm 1)=\lbd\J.\]

We shall begin by examining the automorphisms corresponding to \bref{CC1}. It
follows
from Section~\ref{sec-listing} that \Ut\ may be assumed to be of the form
\begin{equation}\label{G1}
\Ut=\dsum\{\I{a},-\I{b},t\I{c},-t\I{d},\I{a},-\I{b}.
t^{-1}\I{c},-t^{-1}\I{d}\},
\end{equation}
where $a+b+c+d=\ell$.
For any matrix of the form \bref{G1}, we define a function $\theta$, whose
argument is the matrix \Ut\ and whose resultant value is an ordered
quadruplet. Let $\theta$ be specified by
\[\theta(\Ut)=(a,b,c,d).\]
We define \iF{j,k} to be the set containing the automorphism
corresponding to  \Ut\, where \Ut\ is such that
\[\theta(\Ut)=(u_{1},u_{2},u_{3},u_{4})\]
with $(u_{2}+u_{4})=j$ and $(u_{3}-u_{4})=k$. One assumption which may be made
after examining Section~\ref{sec-listing} is that $0\leq u_{2}-u_{4}\leq\elbt$.
We
will now
show that the automorphisms within each set \iF{j,k}\ are mutually conjugate.
To do this, let \Ut\ and \Upt\ be given by \bref{G1}\ such that
\mbox{$\theta(\Ut)=(\np,\nm,\npt,\nmt)$} and \mbox{$\theta(\Upt)=(\np-
1,\nm-1,\npt+1,\nmt+1)$}.
We will now show that the automorphisms \mbox{$\{\Ut,1,\xi\}$} and
\mbox{$\{\Upt,1,\xi'\}$}
are conjugate. Let the matrix \St\ be given by
\[\left(\begin{array}{ccccccc}
\I{n_{+}-1}&\zo&\zo&\zo&\zo&\zo&\zo\\
\zo&\arf(1+t)\I{1}&\zo&\zo&\zo&\frac{i}{2}(1-t)\I{1}&\zo\\
\zo&\zo&\arf(1+t)\I{1}&\zo&\frac{i}{2}(1-t)\I{1}&\zo&\zo\\
\zo&\zo&\zo&\I{\ell-2}&\zo&\zo&\zo\\
\zo&\zo&\frac{i}{2}(1-t^{-1})\I{1}&\zo&\arf(1+t^{-1})\I{1}&\zo&\zo\\
\zo&\frac{i}{2}(1-t^{-1})\I{1}&\zo&\zo&\zo&\arf(1+t^{-1})\I{1}&\zo\\
\zo&\zo&\zo&\zo&\zo&\zo&\I{\ell-n_{+}-1}\end{array}\right).\]
This satisfies
\[\StJSt=\J \; , \; \St\Ut\Stii=\Uppt,\]
where \Uppt\ is some matrix which, according to Section~\ref{sec-listing}, is
conjugate to
the automorphism corresponding to \Upt. Hence the automorphisms
corresponding
to \Ut\ and \Upt\ are conjugate, as we set out to show.

We know already that we may restrict the choice of the parameter $j$ in the
expression \iF{j,k}\ to the values such that $0\leq j\leq\elbt$. In fact it
is possible to restrict the choice of $k$ so that
\begin{equation}\label{G3}
0\leq k\leq(\elbt-j).
\end{equation}
This means that the members of the set \iF{j',k'}\ (where $k'$ does not satisfy
$0\leq k'\leq(\elbt-j')$) are all conjugate to the members of \iF{j,k}\ where
$j,k$ do satisfy \bref{G3}. We prove this by supposing that $j,k$ are such that
they do not satisfy \bref{G3}. We know, however, that the matrix corresponding
to \Ut\ is conjugate to the matrix corresponding to \Utm. A brief inspection
of \Utm\ coupled with results  of Section~\ref{sec-listing} indicate that the
members of \iF{j,k}
are indeed conjugate to the members of \iF{m,n}\, where $m,n$ do satisfy
\bref{G3}.

It follows that every automorphism \bref{CC1}\ belongs to at least one set
\iF{j,k}\ for suitable values of $j,k$. Let us then redefine \iF{j,k}\ slightly
to be the conjugacy class containing the automorphism
corresponding to  \Ut\, where \Ut\ is such that
\[\theta(\Ut)=(u_{1},u_{2},u_{3},u_{4})\]
with $(u_{2}+u_{4})=j$ and $(u_{3}-u_{4})=k$. We have seen that the values
\begin{equation}\label{G4}
j=0,\ldots,\elbt;k=0,\ldots,(\elbt-j)
\end{equation}
are sufficient, in that every automorphism \bref{CC1}\ belongs to one of
the classes \iF{j,k}. We will now show that, for the values specified in
\bref{G4}, the corresponding classes \iF{j,k}\ are all disjoint. Suppose that
\iF{a,b}\ coincides with \iF{c,d}\ for $(a,b)\neq(c,d)$. Then define \Ut\ and
\Upt\ by
\[\begin{array}{cc}
\Ut=\dsum\{\I{n_{+}-b},t\I{b},-\I{a},\I{n_{+}-b},t^{-1}\I{b},-\I{a}\}\\
\Upt=\dsum\{\I{n_{+}-d},t\I{d},-\I{c},\I{n_{+}-d},t^{-1}\I{d},-\I{c}\}.
\end{array}\]
Our supposition implies that for all non-zero $t$
\begin{equation}\label{G5}
\St\ \; \m{U}(st^{\pm 1}) \; \Stii=\lbd t^{\mu}\Upt.
\end{equation}
We know that $s$ may take the values 1 or -1. If $s=1$ and we put $t=1$ in
\bref{G5}\, then we have the necessary condition that $a+c=\ell$, which implies
that
$a=c$. In that case, we put $t=-1$ which implies that $b=d$. This contradicts
our original supposition and so we must have $s=-1$.
However, putting $t=1,-1$ in \bref{G5}\ implies again that $a=c$ and $b=d$.
Thus
the conjugacy classes \iF{j,k}\ where $0\leq j\leq\elbt$ and $0\leq
k\leq(\elbt-
j)$
are disjoint for different choices of $(j,k)$. Moreover, every automorphism
given by \bref{CC1}\ belongs to one of these classes.

Consider next the automorphisms \bref{CC2}\, for which we may assume that
\begin{equation}\label{G6}
\Ut=\dsum\{\I{\ell-a},-\I{a},t\I{\ell-a},-t\I{a}\}.
\end{equation}
As we have mentioned already, it can be shown that $b\leq\elbt$.
Thus, since $a$ can only take values $0,\ldots,\elbt$, we are concerned with
precisely $(1+\elbt)$ distinct automorphisms. Each distinct automorphism
corresponds to a different value of $a$. The automorphisms \bref{CC2}\ fall
therefore into at most $(1+\elbt)$ disjoint conjugacy classes. Let \iG{j}\ be
the conjugacy class containing the automorphism corresponding to \Ut\, where
\Ut\ is given by \bref{G6}\ but with $a=j$. We note that $j$ is restricted
to the values $0,\ldots,\elbt$. We now show that \iG{j}\ is disjoint from
\iG{k}\ when $j\neq k$. Let $\m{U}_{j}(t)$ and $\m{U}_{k}(t)$ be given by
\bref{G6}\ but with $a=j,k$ respectively. If \iG{j}\ and \iG{k}\ are the same,
then
\begin{equation}\label{G7}
\St\;\m{U}_{j}(st^{\pm -1})\;\Stii=\lbd t^{\mu}\m{U}_{k}(t)
\end{equation}
with $s=1,-1$. If we suppose that $s=1$, then putting $t=1$ yields a
contradiction,
since \bref{G7}\ then implies that $j=k$, which is not the case. Thus we
suppose that $s=1$. If we put $t=-1$ in \bref{G7}\, then we again find that
this is not possible. Hence the conjugacy classes \iG{j}\ are disjoint for
distinct values of $j$. Similarly, the conjugacy classes \iF{m,n}\ are disjoint
from the classes \iG{k}. This may be proven by substituting the values
$s=\pm1,t=\pm1$ into a necessary matrix condition. It will be seen that in
both of the cases $s=1,-1$ this proves to be impossible. Clearly, the matrix
equations will have solutions for $\m{S}(\pm1)$, but these will not satisfy
\[\tilde{\m{S}}(\pm1)\;\J\;\m{S}(\pm 1)=\lbd\J.\]
(Much of this has already been done in Section~\ref{sec-1a1} and is applicable
here when dealing with $t$-independent matrices).

Consider the automorphisms \bref{CC3}\, for which we may take
\begin{equation}\label{G8}
\Ut=\dsum\{\I{\ell-a},t\I{a},-\I{\ell-a},-t\I{a}\}.
\end{equation}
We will show that all of these automorphisms are mutually conjugate. Let
\Ut\ and \Upt\ be given by \bref{G8}\ but with $a=j,j+1$ respectively. Then
we define \St\ by
\[\St=\left(\begin{array}{ccccc}
\I{\ell-j-1}&\zo&\zo&\zo&\zo\\
\zo&\arf(1+t)\I{1}&\zo&\frac{i}{2}(1-t)\I{1}&\zo\\
\zo&\zo&\I{\ell-1}&\zo&\zo\\
\zo&\frac{i}{2}(1-t^{-1})\I{1}&\zo&\arf(1+t^{-1})\I{1}&\zo\\
\zo&\zo&\zo&\zo&\I{j+2}\end{array}\right),\]
where $\StJSt=\J$ and $\St\Ut\Stii=\Upt$. Hence all of the automorphisms
\bref{CC3}\ are mutually conjugate. Moreover, this class is disjoint from
all of the other classes identified previously. The method outlined
previously demonstrates that this is so. We call this class \iH.

Consider the automorphisms \bref{CC4}. It follows from
Section~\ref{sec-listing}
that
\Ut\ may be suitably reordered so that
\begin{equation}\label{G9}
\Ut=\dsum\{\I{\ell-a},-\I{a},-t\I{\ell-a},t\I{a}\}.
\end{equation}
We have remarked that the automorphisms corresponding to \Ut\ and $\m{U}(-
t)$
are conjugate. If we then consider $\m{U}(-t)$, where \Ut\ is given by
\bref{G9}\, then it is clear that the automorphisms \bref{CC4}\ belong to the
conjugacy classes \iG{j}\ which we defined earlier.

\end{subsection}
\begin{subsection}{Explicit forms for the automorphisms}
Now we give representatives for the classes which we identified in the previous
subsection.
\begin{enumerate}
\item The class \iF{a,b}\ has the following representative:
\begin{equation}\begin{array}{lll}
\psi(\hk{k})&=&\left\{\begin{array}{l}
				\hk{k}-c \; \mbox{for }k=1; b\neq 0\\
				\hk{k}+c \; \mbox{for }k=b+1; b\neq 0\\
				\hk{k} \; \mbox{otherwise.}\end{array}\right.\\
\psi(e_{j\delta\pm\alpha_{H}})&=&e_{-j\delta\pm\alpha_{H}}\\
\psi(e_{j\delta\pm\alpha_{k}})&=&\left\{\begin{array}{l}
				e_{(-j-1)\delta\pm\alpha_{1}} \; \mbox{for }k=1;
b\neq 0\\
				e_{-j\delta\pm\alpha_{k}} \; \mbox{for }1<k<b+1;
b\neq 0\\
				e_{(1-j)\delta\pm\alpha_{k}} \;\mbox{for } k=b+1;
b\neq 0\\
				-e_{-j\delta\pm\alpha_{k}} \; \mbox{for }k=\ell-a;
a\neq 0\\
				e_{-j\delta\pm\alpha_{k}} \;
\mbox{otherwise}\end{array}\right.\\
\psi(c)&=&-c\\
\psi(d)&=&d-4(\ell+1)(\sum_{p=1}^{b}p\hk{p+1}+b\sum_{p=b+2}^{\ell-1}\hk{p}
+(b/2)\hk{\ell})-2b(\ell+1)c.\end{array}
\label{eq:Frep}\end{equation}
This corresponds to the matrix
\begin{equation}
\dsum\{\I{1},t\I{b},\I{\ell-a-b-1},-\I{a},\I{1},t^{-1}\I{b},\I{\ell-a-b-1},-\I{a}\}.
\end{equation}

\item The class \iG{b}\ has the following representative:
\begin{equation}\begin{array}{lll}
\psi(\hk{k})&=&\left\{\begin{array}{l}
			\hk{k} \; \mbox{for }k=1,\ldots,\ell-1\\
			\hk{\ell}-c \; \mbox{for }k=\ell\end{array}\right.\\
\psi(e_{j\delta\pm\alpha_{H}})&=&e_{-(j+1)\delta\pm\alpha_{H}}\\
\psi(e_{j\delta\pm\alpha_{k}})&=&\left\{\begin{array}{l}
							-e_{-j\delta\pm\alpha_{k}} \;
\mbox{for }k=\ell-b; b\neq 0\\
					e_{-(j+1)\delta\pm\alpha_{\ell}} \; \mbox{for
}k=\ell\\
						e_{-j\delta\pm\alpha_{k}} \;
\mbox{otherwise}\end{array}\right.\\
\psi(c)&=&c\\
\psi(d)&=&d+2(\ell+1)(\sum_{p=1}^{\ell-1}p\hk{p}+(\ell /2\hk{\ell})
+\{(\ell(\ell+1))/2\}c.\end{array}
\label{eq:Grep}\end{equation}
This corresponds to the matrix \mbox{\dsum$\{\I{\ell-b},-\I{b},t\I{\ell-b},-
t\I{b}\}$}.

\item Finally, the class \iH\ has the following representative:
\begin{equation}
\begin{array}{lll}
\psi(\hk{k})&=&\hk{k}\\
\psi(e_{j\delta\pm\alpha_{H}})&=&-e_{-j\delta\pm\alpha_{H}}\\
\psi(e_{j\delta\pm\alpha_{k}})&=&\left\{\begin{array}{l}
				e_{-j\delta\pm\alpha_{k}} \; \mbox{for
}k=1,\ldots,\ell-1\\
				-e_{-j\delta\pm\alpha_{\ell}} \; \mbox{for
}k=\ell\end{array}\right.\\
\psi(c)&=&-c\\
\psi(d)&=&-d.
\end{array}\label{eq:Hrep}\end{equation}
This corresponds to the matrix \mbox{\dsum$\{\I{\ell},-\I{\ell}\}$}.
\end{enumerate}
\end{subsection}
\end{section}

\begin{section} {Summary of conclusions}
\label{sec-summ}

For the complex untwisted affine Kac-Moody algebra \Cli\ the representatives of
the conjugacy
classes of involutive automorphisms are as follows:

\begin{enumerate}
\item Involutive automorphisms of type 1a with $u = 1$:

\begin{enumerate}
\item For the conjugacy classes \iA{j}, where $0\leq j\leq[\frac{\ell}{2}]$,
the representative for \iA{0}\ is the identity automorphism, and the
representatives for \iA{0}\
(for $0< j\leq[\frac{\ell}{2}]$) are given in (\ref{eq:Arep}).
\item For the conjugacy class \iB\ the representative is given in
(\ref{eq:Brep}).
\item For the conjugacy class \iC\ the representative is given in
(\ref{eq:Crep}).
\item For the conjugacy class \iD, which exists only for $\ell$ even, the
representative is given in
(\ref{eq:Drep}).
\end{enumerate}

\item Involutive automorphisms of type 1a with $u = -1$: There is only one
conjugacy class, \iE, for
which the representative is given in (\ref{eq:Erep}).

\item Involutive automorphisms of type 2a (with $u = 1$):
\begin{enumerate}
\item For the conjugacy classes \iF{j,k}, where $0\leq j\leq\elbt$ and $0\leq
k\leq(\elbt-j)$, the
representatives are given in (\ref{eq:Frep}).
\item For the conjugacy classes \iG{j}, where $0\leq j\leq\elbt$, the
representatives
are given in (\ref{eq:Grep}).
\item For the conjugacy class \iH\ the representative is given in
(\ref{eq:Hrep}).

\end{enumerate}

\end{enumerate}
\end{section}

\begin{section}{Acknowledgment}
SPC wishes to acknowledge a research studentship from the SERC.
\end{section}

\end{document}